\documentclass{aa}  

\usepackage{graphicx}
\usepackage{txfonts}
\newcommand{\qdist}[1]{\ifmmode\langle#1\rangle\else\textlangle#1\textrangle\fi}
\begin{document}

   \title{Extra-tidal star candidates in globular clusters of the Sagittarius dwarf spheroidal galaxy\thanks{Tables containig list of extra-tidal star candidates are only available in electronic format the CDS via anonymous ftp to cdsarc.u-strasbg.fr (130.79.128.5) or via http://cdsweb.u-strasbg.fr/cgi-bin/qcat?J/A+A/}}

   \author{Richa Kundu
          \inst{\ref{inst1a}, \ref{inst1}} \thanks{richakundu92@gmail.com}
          \and
          Camila Navarrete
          \inst{\ref{inst1a}, \ref{inst2}}
          \and
          Luca Sbordone
          \inst{\ref{inst1a}}
          \and
          Julio A. Carballo-Bello
          \inst{\ref{insta}}
          \and
          Jos\'{e} G. Fern\'{a}ndez-Trincado
          \inst{\ref{inst3as}}
          \and
          Dante Minniti
          \inst{\ref{inst6}, \ref{inst7}}
          \and
          Harinder P. Singh
          \inst{\ref{inst1}}
          }

   \institute{European Southern Observatory, Alonso de C\'{o}rdova 3107, Vitacura, Santiago, Chile.
              \label{inst1a}
         \and
             Department of Physics and Astrophysics, University of Delhi, Delhi-110007, India.
              \label{inst1}
         \and             
             Millennium Institute of Astrophysics, Av. Vicu\~{n}a Mackenna 4860, 782-0436 Macul, Santiago, Chile.
              \label{inst2}
         \and
             Instituto de Alta Investigaci\'on, Sede Esmeralda, Universidad de Tarapac\'a, Av. Luis Emilio Recabarren 2477, Iquique, Chile.
              \label{insta}
         \and
             Instituto de Astronom\'ia, Universidad Cat\'olica del Norte, Av. Angamos 0610, Antofagasta, Chile
              \label{inst3as}
         \and
             Departamento de Ciencias Fisicas, Facultad de Ciencias Exactas, Universidad Andres Bello, Av. Fernandez Concha 700, Las Condes, Santiago, Chile.
              \label{inst6}
         \and
             Vatican Observatory, V00120 Vatican City State, Italy.
              \label{inst7}
             }

   \date{Received xxxx; accepted yyyy}

 
  \abstract
   {Globular clusters (GCs) associated with the Sagittarius dwarf spheroidal galaxy (Sgr dSph) have evolved under the gravitational potential of both Sgr dSph and the Milky Way. The effects of these potentials are most pronounced in the extra-tidal regions as compared to the central regions of the GCs.} 
   {We aim to study the extra-tidal regions of the GCs that are possibly associated with Sgr dSph, namely Arp 2, Terzan 8, NGC 5634, NGC 6284, Terzan 7, NGC 2419, NGC 4147, M 54 and Pal 12, using data from the {\it Gaia} early data release 3.}
   {We selected the extra-tidal candidates based on their angular distances from the cluster centre in the RA-Dec plane, proper motions of the clusters and the individual extra-tidal star candidates, and their positions on the colour-magnitude diagrams of the clusters.}
   {We found extra-tidal candidates for the nine studied GCs. For eight of them, the surface density of candidate extra-tidal stars in the vicinity of the clusters is in significant excess with respect to more distant surrounding fields. No extended extra-tidal features beyond 5 tidal radii were detected for any of the clusters.}
   {We publish a list of the most probable extra-tidal candidates that we determined using {\it Gaia} astrometric and photometric data. Our analysis shows that the clusters that are associated with Sgr dSph are more likely affected by the gravitational potential of the Sgr, as the distribution of extra-tidal stars is elongated in the same direction as the local stream. NGC 4147 is the only exception. We found some high-probability candidate extra-tidal stars in several of the analysed clusters. We failed to detect any coherent large-scale tidal tail around them.}

\titlerunning{Extra-tidal star candidates in Sgr dSph Galaxy}
\authorrunning{Kundu et al.}



   \maketitle
%

\section{Introduction}

In the $\Lambda$ cold dark matter universe, the growth of galaxies is hierarchical, meaning that dwarf galaxies are cannibalised by large galaxies \citep{Gomez16, Bullock17r}. The Sagittarius dwarf spheroidal galaxy (Sgr dSph) \citep{Ibata94, ibata95} is a clear example of such a phenomenon. Soon after its discovery, the first numerical simulations predicted that  Sgr dSph is undergoing tidal disruption \citep{Johnston95} due to its interaction with the Milky Way (MW). Indeed, thanks to all-sky and pencil-beam surveys, several clear detections of the stellar stream were achieved across the whole sky \citep[e.g.][]{Delgado02, Newberg02, Majewski03, Belokurov06, Koposov12, Huxor15, Navarrete17, Hasselquist19, Hayes20, antoja20, ibata20}. Sgr dSph thus represents an excellent laboratory to study the past, present, and future dynamic states of stellar systems subject to ongoing tidal stripping. In this context, while its main body helps us to understand the state of the galaxy before the stripping process took place, the outer regions of the core and the remarkable streams provide information about the present and future stages, respectively \citep{lawbook}.

A number of globular clusters (GCs) are believed to have formed within Sgr dSph, with varying degrees of certainty \citep[see e.g. ][and references therein]{Massari19, antoja20, forbes20, Bellazzini20, dante21}. Four of them, namely M 54 (NGC 6715), Terzan 7, Terzan 8, and Arp 2, are in close proximity or within the main body. They share distance, radial velocity (RV), and proper motion (PM) with this main body, and are thus believed to be still gravitationally bound to it \citep[see e.g. ][and references therein]{Costa95, law10, Sohn18, Bellazzini20}. In particular, M 54 was found to reside in the Sgr dSph core by \citet{monaco05} when they analysed the density profile of the galaxy, although it is possible that it formed outside the core and  later fell in the central potential well, perhaps as a result of the interaction with the MW \citep{Bellazzini08}. Based on the high-resolution analysis of five giant stars from Terzan 7, \citet{luca05} found that the cluster shares the same low $\alpha$/Fe and low Ni/Fe ratios as Sgr dSph stars in the same metallicity range, confirming its association.

Many other MW clusters, currently not close to Sgr dSph, show (to varying degrees of confidence) kinematical, positional, or chemical indications of having formed in that galaxy, and having been subsequently stripped. Among them, the best case is Palomar 12 (Pal 12), which is considered as being associated with the Sgr stream on kinematical grounds \citep{martnez02, Bellazzini03, law10, Sohn18}. It also shares a number of highly peculiar chemical signatures with Sgr dSph \citep{cohen04, luca07} and is thus considered a bonafide Sgr dSph member. 

NGC 4147 has been proposed as a candidate member by some authors \citep{Bellazzini03, law10}, although recent studies have discarded any association with Sgr dSph by means of PM analysis and chemical abundances \citep[see e.g. ][]{Villanova16, Sohn18}. A possible connection between the cluster and one of the oldest arms of the Sgr stream has been recently proposed by \citet{Bellazzini20}, based on RR Lyrae stars. High-resolution abundance analyses of NGC 5053 and NGC 5634 \citep{luca15, Carretta17} show plausibly but inconclusively that they might have originated in Sgr dSph, while \citet{tang18} discard any association between NGC 5053 and Sgr dSph based on tentative chemical evidence (mostly different [Mg/Fe] abundances) and the past orbit of the cluster. Recent studies have associated the cluster with the LMS-1 stellar stream \citep{Yuan20, Malhan21}. The kinematic study of Withing 1 stars by \citet{Julio17} show that the velocity components of the cluster match with the velocity component of leading and trailing arms of the Sgr stream, supporting the association between this young \citep[$5$ Gyr, ][]{Carraro07} GC and Sgr dSph. 

From the \citet[][hereafter LM10]{law10b} analysis, NGC 2419 was not considered to be associated with the Sgr stream due to its large distance. However, \citet{Belokurov14} demonstrated that this cluster, the most distant GC from the Sgr dSph centre, is located at the apo-centre of the trailing arm of the stream. This association was recently confirmed, based on its kinematics and age, by \citet{Massari19}, along with Withing 1 and Pal 12.

The distribution of stars in the extra-tidal region of a GC is affected by various forces acting on it. The GCs associated with Sgr dSph have evolved under the potential of both MW and Sgr dSph itself. Therefore, by studying the outer regions of these clusters, we can gain insights into the prominent forces dominating the various regions of the Sgr dSph system. We aim to study the extra-tidal area around these clusters in a systematic way, using both astrometry and photometry to assess the level and distribution of possible extra-tidal stars on top of Sgr dSph and Sgr stream stars. To the best of our knowledge, no systematic study of the extra-tidal regions of these Sgr's GCs has been performed, although some clusters have been individually studied searching for tidal tails, using different techniques and data sets \citep[e.g. ][]{leon00, Julio14, Jordi10}.

The European Space Agency's {\it Gaia} mission \citep{gaiadr1} observations have been very helpful in the studies of MW tidal streams \citep{Ibata18, Malhan18b, Ibata19a, Ibata19b, Palau19, antoja20}, providing precise PMs for the first time, along with parallaxes for more than one billion sources. The {\it Gaia} early data release 3 \citep[][hereafter {\it Gaia} EDR3]{gaiadr3} published precise PMs (with errors of the order of 0.05 mas yr$^{-1}$ for sources brighter than G = 17 mag) for an unprecedentedly large number (over 1.4 billion) of stars. We used {\it Gaia} EDR3 astrometry and photometry data to study the extra-tidal regions of 11 GCs, based on the recent results of \citet[][hereafter B20]{Bellazzini20}: M 54, Terzan 7, Arp 2, Terzan 8, Pal 12, Whiting 1 (confirmed members), NGC 2419, NGC 5634, NGC 4147 (likely to be associated with Sgr dSph), Pal 2 and NGC 6284 (unlikely candidate members). Recently, \citet{dante21, dante21a} discovered 20 new GC candidate members of  Sgr dSph. As the nature of some of these low-luminosity cluster candidates \citep{Garro21} embedded in the main body of Sgr is still under debate \citep{piatti21}, they are not included in the present work.

This paper is organised as follows: in Section 2 we describe the methodology followed to select the extra-tidal candidates for the clusters; in Section 3 we estimate the level of contamination around the clusters, along with the possible directions of any extended extra-tidal features that may be present; in Section 4, we search for RV measurements in the literature of the extra-tidal star candidates to confirm their origin; and in the last two sections, we present our results and our concluding remarks.

\section{Selecting extra-tidal candidates}

\begin{table*}[htbp]
        \centering
        \caption{Physical and kinematic parameters of the clusters studied in this work.}
        \label{tab:par}
        \begin{tabular}{cccccccccc} 
\hline

Cluster name & RA       & Dec & $r_t^{\dagger}$ &  $r_J^{\ddag}$ & $\mu_{\alpha}cos\delta$ & {${\sigma_{\mu}}_{\alpha}$} & {$\mu_{\delta}$} & {${\sigma_{\mu}}_{\delta}$} & Correlation\\
           & (deg)   &  (deg) & (arcmin) &  (arcmin) & (mas yr$^{-1}$)  & (mas yr$^{-1}$) & (mas yr$^{-1}$)  & (mas yr$^{-1}$) & \\
\hline
Arp 2    & 292.184 & --30.356 & 4.61$^{\beta}$ & 9.24$^*$ & --2.37 & 0.42 & --1.51 & 0.25 &   0.57\\
Terzan 8 & 295.435 & --33.999 & 3.96           & 8.88     & --2.48 & 0.24 & --1.58 & 0.17 & --0.59\\
M 54     & 283.764 & --30.480 & 7.56           & 39.30    & --2.68 & 0.29 & --1.39 & 0.22 &   0.56\\
Terzan 7 & 289.433 & --34.658 & 7.38           & 8.76$^*$ & --2.97 & 0.22 & --1.65 & 0.17 &   0.35\\
NGC 5634 & 217.405 & --5.976  & 8.40           & 22.08    & --1.70 & 0.23 & --1.47 & 0.19 & --0.90\\
NGC 2419 & 114.537 &  38.882  & 7.50           & 30.12    & --0.04 & 0.47 & --0.52 & 0.29 &   0.70\\
NGC 4147 & 182.525 &  18.542  & 6.06           & 18.18    & --1.71 & 0.21 & --2.08 & 0.19 & --0.75\\
Pal 12   & 326.662 & --21.253 & 13.20$^!$      & 8.88     & --3.22 & 0.34 & --3.36 & 0.17 &   0.62\\ 
NGC 6284 & 256.121 & --24.764 & 3.69$^{\ddag}$ & 19.80    & --3.19 & 0.34 & --2.02 & 0.15 &   0.40\\
\hline
\end{tabular}
\tablefoot{$\dagger$:$r_t$ from \citet{mackey05} and converted to arcmin using the distances from \citet{harris96}; $\ddag$: from \citet{Boer19}; $\beta$: \citet{Salinas12}; $!$: \citet{Musella18}; $*$: \citet{Baumgardt10}.}
\end{table*}

\begin{table*}
        \centering
        \caption{Parameters of the clusters used to generate the isochrones and ZAHBs.}
        \label{tab:iso}
        \begin{tabular}{ccccccc} 
\hline

Cluster name &  [Fe/H]$^{\P}$ & Age$^{\P}$ &    Distance$^{\S}$ & Distance (adopted) & [$\alpha$/Fe]$^\&$ & Initial mass on HB$^@$\\
           & (dex) & (Gyr) & (kpc) & (kpc) & & (M$_{\odot}$) \\
\hline
Arp 2     & -1.45 & 10.9 & 28.6 & 28.6      & 0.31$^a$  & 0.80\\
Terzan 8  & -1.80 & 12.2 & 26.7 & 30.0      & 0.37$^a$  & 0.80\\
M 54      & -1.25 & 10.7 & 26.7$^*$ & 26.7  & 0.21$^b$  & 0.85\\
Terzan 7  & -0.56 & 7.3  & 22.8 & 22.8      & -0.03$^c$ & 1.00\\
NGC 5634  & -1.94 & 11.8 & 27.2 & 27.2      & 0.20$^d$  & 0.80\\
NGC 2419  & -2.14 & 12.3 & 83.2 & 100.2     & 0.13$^e$  & 0.80\\
NGC 4147  & -1.50 & 12.0 & 18.2 & 19.9      & 0.38$^f$  & 0.85\\
Pal 12    & -0.83 & 8.8  & 19.0 & 19.0      & -0.20$^g$ & 0.90\\
NGC 6284  & -1.13 & 11.1 & 15.1 & 15.1      & 0.47$^h$  & 0.90\\
\hline
\end{tabular}
\tablefoot{$\P$: from \citet{Forbes10}; $\S$: from \citet{Baumgardt19}; $*$: \citet{Hamanowicz16}; $\&$: the allowed range of [$\alpha$/Fe] is between 0.0 and 0.3, when the literature value is out of the allowed range, the value closest to the range is selected;$@$: \citet{Valcarce12}; $a$: \citet{Mottini08}; $b$: \citet{Brown1999}; $c$: \citet{luca03}; $d$: \citet{Carretta17}; $e$: \citet{Sharina13}; $f$: \citet{Villanova16}; $g$: \citet{brown95}; $h$: \citet{Puzia05}.} 
\end{table*}

In this work, astrometry and photometry data from {\it Gaia} EDR3 was used. We first selected a sub-sample of astrometrically well-behaved sources, following the same criteria suggested in the data release papers \citep{gaiadr3, antoja21}:\\
\texttt{1.) RUWE $<$ 1.4}\\
RUWE is the re-normalised unit weight error. This condition ensures that we use the stars whose astronomical observations are well fitted by a single-star model. Higher values may indicate some problematic or non-single sources.\\
\texttt{2.) ASTROMETRIC\_EXCESS\_NOISE\_SIG $\leq$ 2.}\\
Excess noise is the extra noise in each observation assumed to explain the residual scatter in the astrometric solution. If ASTROMETRIC\_EXCESS\_NOISE\_SIG is greater than two, then this excess noise is statistically significant.\\
\texttt{3.) ASTROMETRIC\_GOF\_AL $<$ 3}\\     
This parameter represents how good the fit is between the astrometric model and the observations. Higher values indicate a bad fit.\\
\texttt{4.) VISIBILITY\_PERIODS\_USED > 10}\\ 
This parameter indicates the set of observations separated by at least 10 days. A higher value indicates that the source is well observed.\\     
\texttt{5.) 0.001+0.039(BP-RP) < log10 excess\_flux < 0.12+0.039(BP-RP)}\\
Excess\_flux is the corrected phot\_bp\_rp\_excess\_factor \citep{Riello21}. This factor can be estimated using the Python code provided by \citet[][Fig. B.2]{gaiadr3}. The sources that are out of this range have inconsistent fluxes for various reasons, such as the presence of another nearby source, or the observed source may be an extended source.\\

We applied the above criteria to all the stars within five times the \citet{King62} tidal radii ($r_t$) of the nine clusters. Two additional clusters, Pal 2 and Withing 1 were originally part of the sample, but they did not have enough astrometrically well-behaved stars (310 and 13 sources inside their $r_t$ for Pal 2 and Withing 1, respectively) to get a clean colour-magnitude diagram (CMD) of the cluster stars and were therefore excluded.

In order to select the extra-tidal candidates of each cluster, we select the stars with similar PMs and CMDs as the cluster stars inside one $r_t$, similar to the criteria used in \citet{kundu19, kundu19b, Kundu20}. Stars lying between $r_t$ to 5$r_t$  from the cluster centre were considered as the extra-tidal candidates in this study. The coordinates, adopted $r_t$, and Jacobi radius ($r_J$) \footnote{The stars that are outside the $r_J$ are completely out of the gravitational potential of the cluster. However, the stars that are inside the $r_J$ but outside the $r_t$ are potential escapers, and may still be bound to the cluster. See e.g. \citet{Fukushige2000, Baumgardt10, Kupper10,  julio11, Claydon10}.} of each cluster are listed in Table~\ref{tab:par}.
\\

\subsection{Proper motion selection}\label{sec:ppm}

\begin{figure*}
\begin{center}
\includegraphics[width=\textwidth, height=\textheight, keepaspectratio]{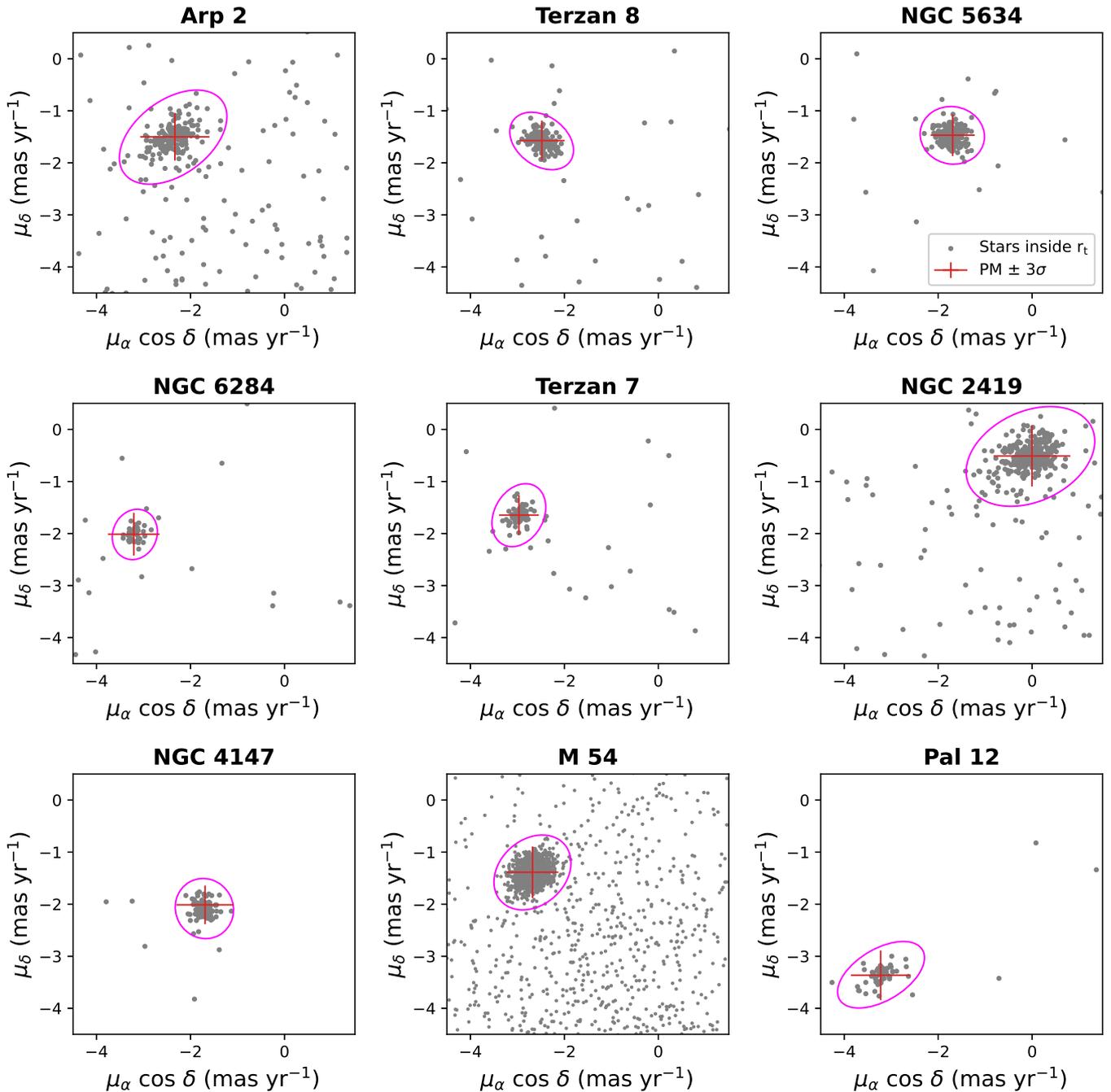}
\caption{Centre of the PM distribution $\pm$ 3$\sigma$ (red lines) for each cluster, determined as the centre and dispersion of the Gaussian fit of the cluster stars inside one $r_t$ (grey dots). For Arp 2, Terzan 7, NGC 6284 and Pal 12, one third of the nominal $r_t$ was used (see the text for further details).}
\label{PM}
\end{center}
\end{figure*}

Using the sub-sample of astrometrically well-behaved sources inside their $r_t$, we first estimate the PM distributions of each of the nine clusters. To do this, we used a Gaussian mixture model, including two components: one narrow Gaussian distribution for the cluster stars, and one broad Gaussian distribution for the field population. In order to fit the models, the Python package \texttt{Scikit-Learn} was used \citep{sklearn}, which finds the best model to fit the data using an expectation-maximisation algorithm. A full covariance was used, allowing  asymmetrical distributions to be fitted. From the best fit, the mean and standard deviation for the cluster and field population were recovered. The linear correlation coefficient, $\rho$, was derived from the resulting (non-diagonal) correlation matrix, calculating the tangent of the eigenvalues of the covariance matrix. The central value of the prominent Gaussian distribution is taken as the PM of the cluster, and the sigma of the distribution, $\sigma_{\rm cluster}$, is considered as the intrinsic dispersion of the cluster PM distribution. Figure~\ref{PM} shows the PM vector-point diagram for each cluster, showing the one $\sigma$ ellipse for the best fit model PM distribution. In the case of Arp 2, Terzan 7, NGC 6284 and Pal 12, only stars brighter than G = 19 mag and inside one-third of $r_t$ were used for the figure and the Gaussian fit, as the clusters are embedded in high-density regions that dominate the PM diagram over the scarce cluster population when all the sources inside the nominal $r_t$ are considered. The estimated PM values, along with the dispersion in each direction and the linear correlation coefficient for each of the clusters, are listed in Table~\ref{tab:par}. Tables~\ref{tab:ppm} and~\ref{tab:ppm_field} present the uncertainties associated with each measurement, including the parameters for the field population around each cluster region.

From Fig.~\ref{PM} it can be seen that, for most of the clusters, the PMs are not completely symmetrical and their dispersions can be significant. Therefore, in order to include all possible extra-tidal candidates, we need to search for stars that have compatible PMs, and
take into consideration the dispersion in the PM distributions. This is the reason why we decided to perform a fit to the PM distribution instead of using the values already reported in the literature. In fact, our adopted PM values agree with the values reported in \citet{GC}, within the errors. However, the errors published in the aforementioned study do not represent the intrinsic dispersion on the PM distribution, but instead the error on the mean, and a selection based on these values would lead to the loss of many probable extra-tidal candidates. \cite{GC} published the PM dispersion profile for a sub-sample of clusters, including M 54, NGC 5634, NGC 4147 and NGC 6284, while the rest of the clusters studied in this work are not included. We compared these values with those derived in this work, for the four clusters in Appendix~\ref{sec:appendix}.

As the clusters are embedded inside the Sgr stream stars, the PM distribution could have been fitted with a model including a superposition by three Gaussians. However, the Sgr stream stars' PMs are almost indistinguishable from the PM distribution of the cluster, which is the most prominent inside one $r_t$. To decide the number of Gaussians to be used in the fit, we used the Bayesian Information Criterion \citep[BIC,][]{Schwarz78} and Akaike Information Criterion \citep[AIC,][]{Akaike74}, finding that there was no significant difference between the information criteria values obtained when using two or three Gaussian components. Repeating the fit to the data, through the expectation-maximisation algorithm, a narrow distribution for the cluster stars, and a broad distribution for the field stars were always recovered, while the third component's mean and standard deviation change from one realisation to the next. Therefore, we decided to use a bivariate Gaussian fit considering only two Gaussians, at the expense of including Sgr core and stream stars in the fit for the Gaussian distribution of the cluster stars.

Given the mean and the dispersion of the PM distribution of the cluster stars, we proceed to estimate the probability of each star (up to five $r_t$) to be associated with a given cluster, based on its PM. To estimate the probability, we tested the two following methods: (1) In the first method, for each star, the PM in the RA and Dec planes was modelled as a bivariate function, with mean value as the PM of the star, and the covariance matrix containing sigma errors and correlation in the PM. To estimate the probabilities, we integrated the bivariate function between the range of the PM defined by the cluster population (PM$_{\rm cluster}$ $\pm$ 3$\sigma_{\rm cluster}$; seventh and ninth columns in Table~\ref{tab:par}); (2) In the second method, membership probabilities were computed adopting the same formulation and methodology as described in Section 4 of \citet{Sariya15}. Here, we adopted the mean PM of the cluster from estimates in the previous section and assumed the radius of the distribution as the intrinsic dispersion on the cluster's PM. The errors and the correlation between the two PM components used to estimate the probabilities were also determined using the member stars (the stars which are inside the $r_t$ of the cluster).\\

To select one of the above approaches, we tested both of them using RV data available in the literature. \citet{Kimmig15} recently published RV data for 25 GCs. Two clusters, namely M 54 and NGC 4147, were common between their sample and ours. M 54 has a wide range of RVs \citep[between 100 km s$^{-1}$ and 180 km s$^{-1}$, ][]{Bellazzini08}, which could lead to the inclusion of many false positives. Therefore, it does not present a good testing environment. Hence, we tested these approaches with NGC 4147. We had 14 stars in our search radius and five of them had RVs compatible with the cluster \citep[183.2$\pm$0.7 km s$^{-1}$, ][]{harris10}, within 3$\sigma$ error. We found that, based on the first approach, all 14 stars were selected as part of the cluster, with probabilities $>$ 50\%. However, based on the second approach, just three of the 14 stars had probabilities of being part of the cluster $>$ 50\% and were selected based on their PMs. These three stars had RVs compatible with the cluster. According to this simple test, in the case of NGC 4147, we found that the first approach provides a complete list of extra-tidal candidates but at the expense of including some false positives (in this case 64\% false positives). However, in the second approach, the list of the extra-tidal candidates may not be complete but it contains the most probable extra-tidal candidates. In this paper, we adopted the second approach to analyse the extra-tidal regions of the clusters.

To select the extra-tidal candidates, first, the probabilities of the stars having compatible PM with the cluster were estimated. The PM of the clusters (especially in core clusters) is similar to the Sgr field stars. This could lead to the misclassification of Sgr field stars as  cluster stars. Hence, we also calculated the probability of the stars being compatible with the PM of the Sgr dSph field population. To get the PM of the Sgr dSph field population, Eqn (1) from \citet{ibata20} was used, converting the PMs from the heliocentric Sagittarius coordinates into (RA, Dec) celestial coordinates using the Gala python module \citep{galacoor}. The stars from the main body of Sgr dSph generally have PMs that are similar to those of the inner clusters due to their dynamical common origin, and hence the Eqn (1) from \citet{ibata20} is helpful to separate both populations. Eqn 1 in \citet{ibata20} is a linear fit to the PM in the Lambda direction, assuming that the Sgr stars are moving in a plane. Thus, we adopted $\mu_B+\mu_{\rm B, reflex}=0.0$ mas yr$^{-1}$. We adopted the standard deviation on the mean PMs as 0.4 mas yr$^{-1}$, also mentioned by \citet{ibata20}. Then, only the stars whose total probability of having a PM compatible with the PMs derived for the cluster was greater than 50\%, but also whose probability was equal to or greater than the probability of having a PM compatible with the Sgr dSph field population, were selected and further analysed in the next step.

\subsection{Colour-magnitude selection}
The CMDs of the cluster stars were constructed using the {\it Gaia} EDR3 photometry. To select the cluster stars, only the stars lying inside the $r_t$ of the cluster and having similar PMs as derived for the cluster were considered. Up to this point, the individual probabilities based on the PMs were not considered (for cluster stars), as the goal was to recover the CMD of the cluster, even if some faint cluster stars tend to have lower probabilities (due to large errors in PMs). The CMDs of the clusters were de-reddened using the E(B-V) value for each star from the \citet{Schlafly11}\footnote{\url{https://irsa.ipac.caltech.edu/applications/DUST/}} reddening maps and the corresponding coefficients for the {\it Gaia} band-passes (retrieved from the PARSEC\footnote{\url{http://stev.oapd.inaf.it/cmd}} web interface). From the CMD of the clusters, we traced the cluster population with the PARSEC stellar isochrones \citep{Bressan12, Marigo17}, while the horizontal branch (HB) populations were traced using the zero-age horizontal branch (ZAHB) models\footnote{\url{http://www2.astro.puc.cl/pgpuc/zahbs.php}} from \citet{zahb}, adopting the stellar parameters (distance, metallicity, [$\alpha$/Fe], and initial mass on the HB) from various sources in the literature, as listed in Table~\ref{tab:iso}.\\

\begin{figure*}
\begin{center}
\includegraphics[width=0.94\textwidth, height=0.94\textheight, keepaspectratio]{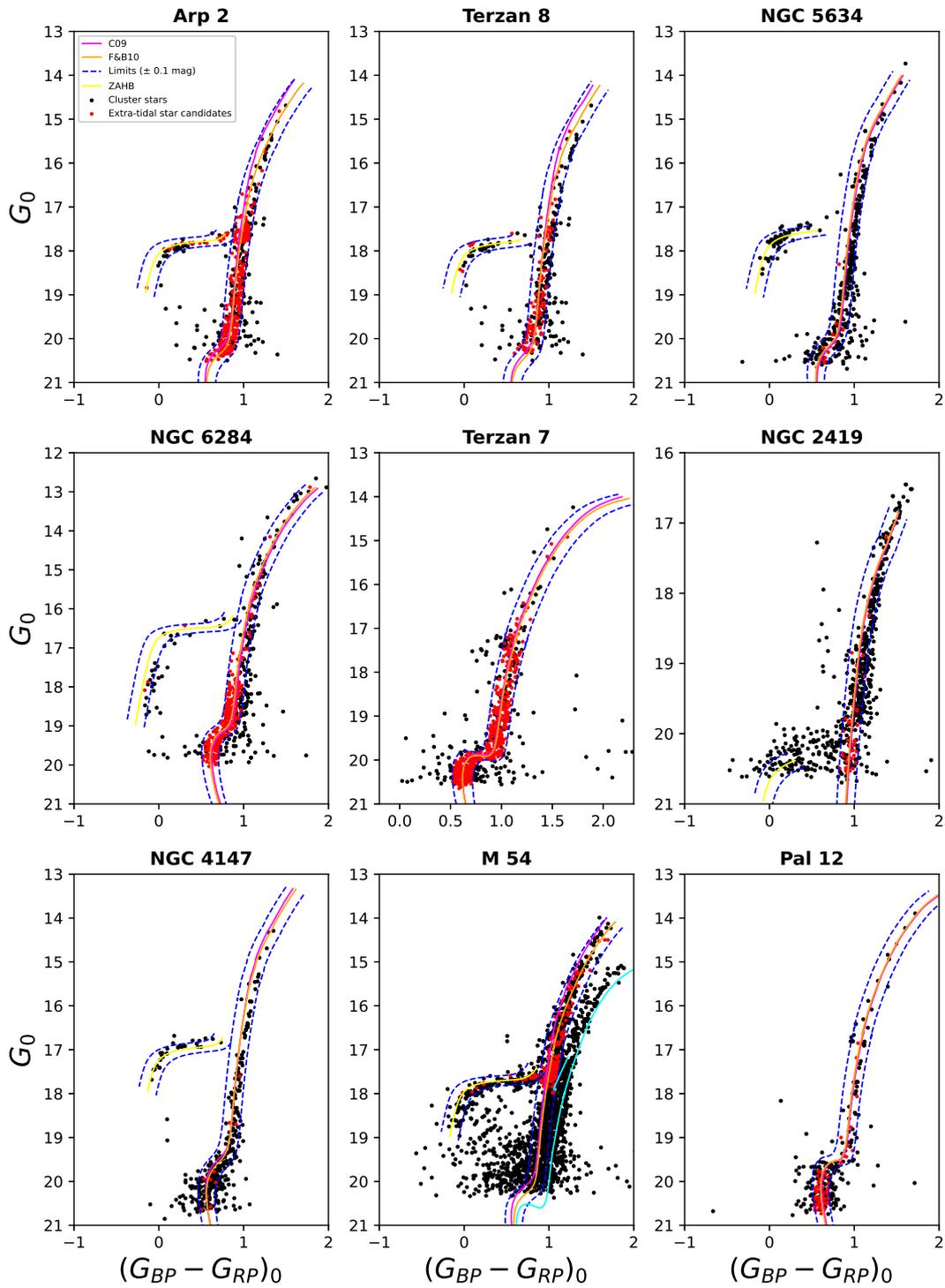}
\caption{{\it Gaia} EDR3 de-reddened CMDs of the clusters (grey dots) along with the selected extra-tidal candidates (red dots) and PARSEC isochrones (solid lines). The orange and pink lines represent the isochrones for each cluster based on parameters from the literature (see the text for further details). The yellow line corresponds to the ZAHB ridgeline at the distance of each cluster. The dashed blue lines define the selection cuts applied in order to select the extra-tidal candidates around the clusters.}
\label{cmd}
\end{center}
\end{figure*}

Figure~\ref{cmd} shows the isochrones (orange and pink lines), along with the ZAHBs (yellow line) and the {\it Gaia} EDR3 de-reddened CMDs of the clusters. Grey dots represent the cluster stars whose PMs match with the PMs of the clusters within a 1$\sigma$ range. Black dots represent the cluster stars whose probability of having a PM compatible with the cluster is more than 50\% and also equal to or more than the probability of having a PM compatible with the Sgr dSph population. The orange line represents the isochrone based on the adopted parameters \citep{Forbes10}, while the pink line shows isochrones  based on the parameters from \citet{Carretta09}, except for Terzan 7 \citep[from][]{Sbordone07} and Terzan 8 \citep[from][]{harris10}. It can be seen that the CMDs are in good agreement with the isochrones based on the \citet{Forbes10} stellar parameters, although part of the red giant branch (RGB) tends to be slightly redder than the isochrone. The blue lines show the limits ($\pm$ 0.1 mag in colour and G-magnitude) adopted in this work for selecting a star as an extra-tidal star for the cluster. Finally, the extra-tidal candidates selected based on these limits are shown with red dots. Two stellar sequences can be seen in the case of M 54, both having similar PMs, inside one $r_t$. The second, reddest sequence, belongs to a much younger population from Sgr dSph, which is traced by the light blue isochrone in Figure~\ref{cmd}, corresponding to the well known high-metallicity population in the core of Sgr dSph \citep{Bonifacio04, Sbordone07, Siegel07}. Both RGBs, from the cluster and the Sgr dSph population, could overlap around G $\sim$ 18 mag. Therefore, to avoid including contamination from Sgr dSph, the selection of potential extra-tidal candidates in this case was restricted to stars brighter than G = 18 mag.
\\

The mean parallax and corresponding dispersion for each cluster were calculated using the cluster stars. The distribution of some clusters includes also negative parallaxes, which are indicative of the uncertainty on the parallax measurement and the distribution itself \citep[see e.g. ][]{Pancino17}. However, \citet{Lindegren18} concluded that the negative parallaxes are a natural outcome of {\it Gaia} measurements. Therefore, extra-tidal star candidates whose parallaxes were out of the 3$\sigma$ range of the parallax distribution were discarded, while the stars with parallaxes compatible with the cluster parallax distribution were kept, even if the values were negative. To discard nearby stars, sources with parallax values greater than 0.5 mas were discarded from the extra-tidal star candidates.
\\

\begin{figure*}
\begin{center}
\includegraphics[width=\textwidth, height=\textheight, keepaspectratio]{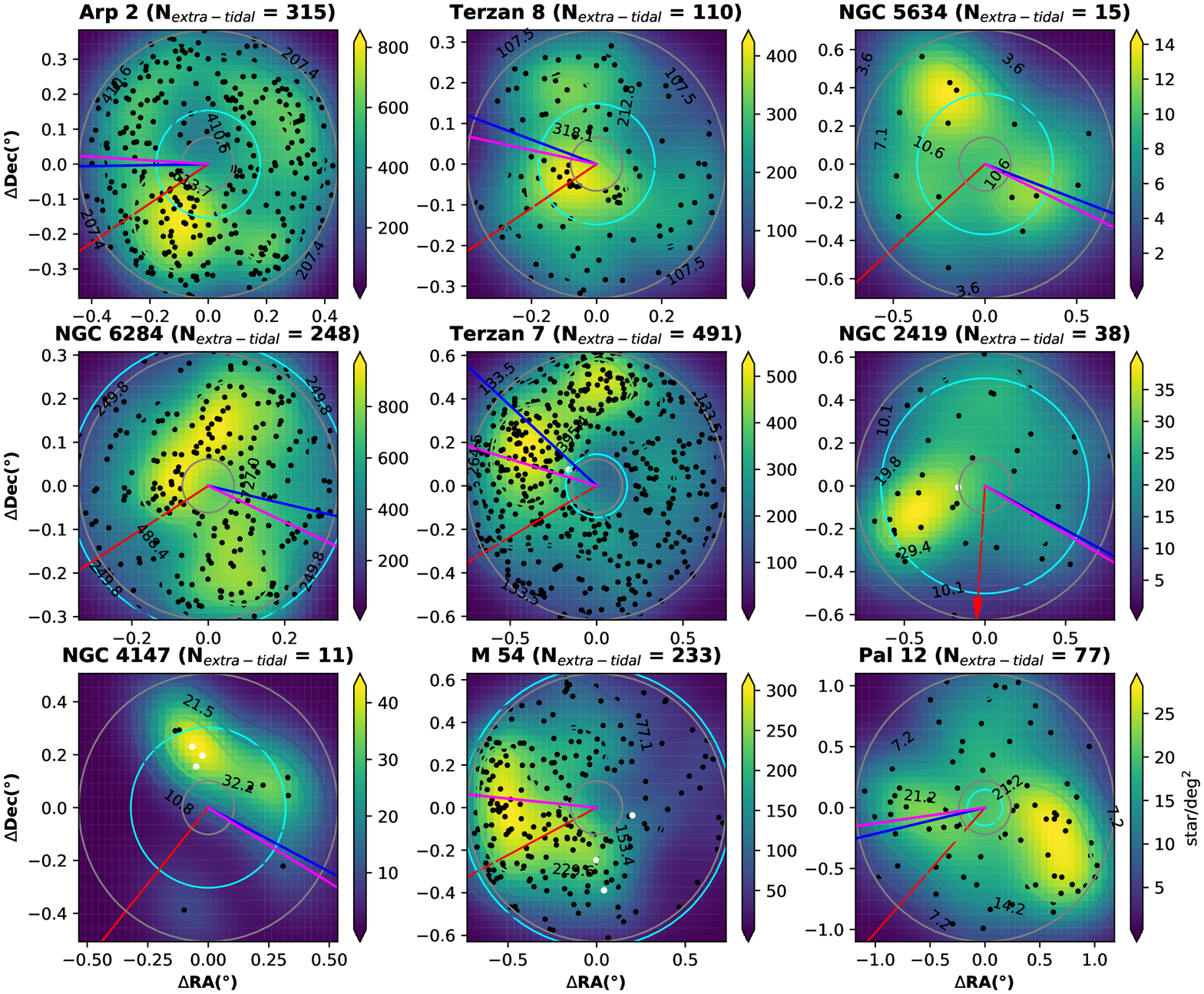}
\caption{Density map of the selected extra-tidal cluster candidates (black dots), along with the PMs (red), $r_t$ (grey), 5$r_t$, and $r_J$ (light blue). The directions of the Galactic centre (pink) and to the core of Sgr dSph (blue) are also shown. Contour lines represent the iso-density regions with a constant number of stars per square degree}
\label{map}
\end{center}
\end{figure*}

The density maps of the selected extra-tidal candidates (N$_{extra-tidal}$), along with the PM direction (red line), $r_t$ and $r_J$ of each cluster are shown in Figure~\ref{map}. The contour lines represent the iso-density regions with a constant number of stars per square degree. In the title of each panel, the name of the cluster and the number of extra-tidal star candidates are indicated. The density maps were created using the kernel density estimator (KDE) routine from the \texttt{scipy} module \citep{scipy}. The bandwidths of the Gaussian KDE used were 5.0, 5.0, 12.8, 3.8, 8.0, 12.7, 7.43, 8.3, and 16.8 arcmin for Arp 2, Terzan 8, NGC 5634, NGC 6284, Terzan 7, NGC 2419, NGC 4147, M 54, and Pal 12, respectively. The figure also shows the direction towards the Galactic centre (pink) and towards the core of Sgr dSph (blue). The light blue circle,  the inner grey circle, and the outer grey circle correspond to the $r_J$,  one $r_t$, and five $r_t$, respectively. Finally, to have a better understanding of these extra-tidal candidates\footnote{The full list of the stars considered in each cluster with their classification will be made available as online material.}, we estimated their significance over the field population in the next section.

\section{Contamination Analysis}

In this section, we estimate the amount of possible contaminations due to the Galaxy and Sgr dSph populations in the number of extra-tidal candidates that we estimated in the previous section. To do this, we selected several regions around each GC. These regions are at least 10$r_t$ away from the cluster centre and each has an area of 5$r_t$. We cleaned the samples to retain only astrometrically well-behaved sources and then selected the stars by applying the same criteria as explained in Section 2. We do not consider any variation or gradient in the PMs of the Sgr dSph population because our aim is to estimate the number of Sgr or MW stars that randomly fall inside the same selection cuts, as the extra-tidal candidates.

In the contamination analysis, the clusters closest to the Sgr dSph main body were handled very carefully as the density of the field stars around them changes rapidly with a slight change in the coordinates. Hence, we used data from B20 to select as comparison fields those having a similar stellar density to that of the cluster (Bellazzini, M., private communication). M 54 was not considered as it coincides with the Sgr dSph core and, therefore, we could not find any other region with a similar number density population to that of the cluster. In the case of Arp 2, Terzan 7, and Terzan 8 (the closest to the core of Sgr), we selected only two comparison fields, which approximately lie on the same iso-density contours of the Sgr dSph population as that of the clusters. The two comparison regions are located at an angular separation of 15, 21, and 34, and 18, 34, and 39 times $r_t$ for Arp 2, Terzan 7, and Terzan 8, respectively. This angular separation goes from 2.2 degrees to 3.8 degrees. The CMDs and vector point diagrams for these clusters and the two comparison fields are included in Appendix~\ref{sec:appendix_fields}.

For the other clusters that are more distant from the core of Sgr dSph, namely NGC 2419, NGC 4147, NGC 6284 and Pal 12, we selected eight comparison fields, in eight different directions around the clusters. These eight fields were in the northern, southern, eastern, western, north-eastern, north-western, south-eastern, and south-western directions from the cluster centres. The direction of increasing Dec was considered as north. For each of the clusters, northern, southern, eastern and western regions were 11$r_t$, and north-eastern, north-western, south-eastern and south-western regions were 15.5$r_t$  from the corresponding cluster centre. Having several fields in different directions gave us the possibility of detecting the presence of any possible extended extra-tidal features beyond five times the $r_t$, in a particular direction, if an overdensity of likely cluster star was found.
\\

\begin{figure*}
\begin{center}
\includegraphics[width=\textwidth, height=\textheight, keepaspectratio]{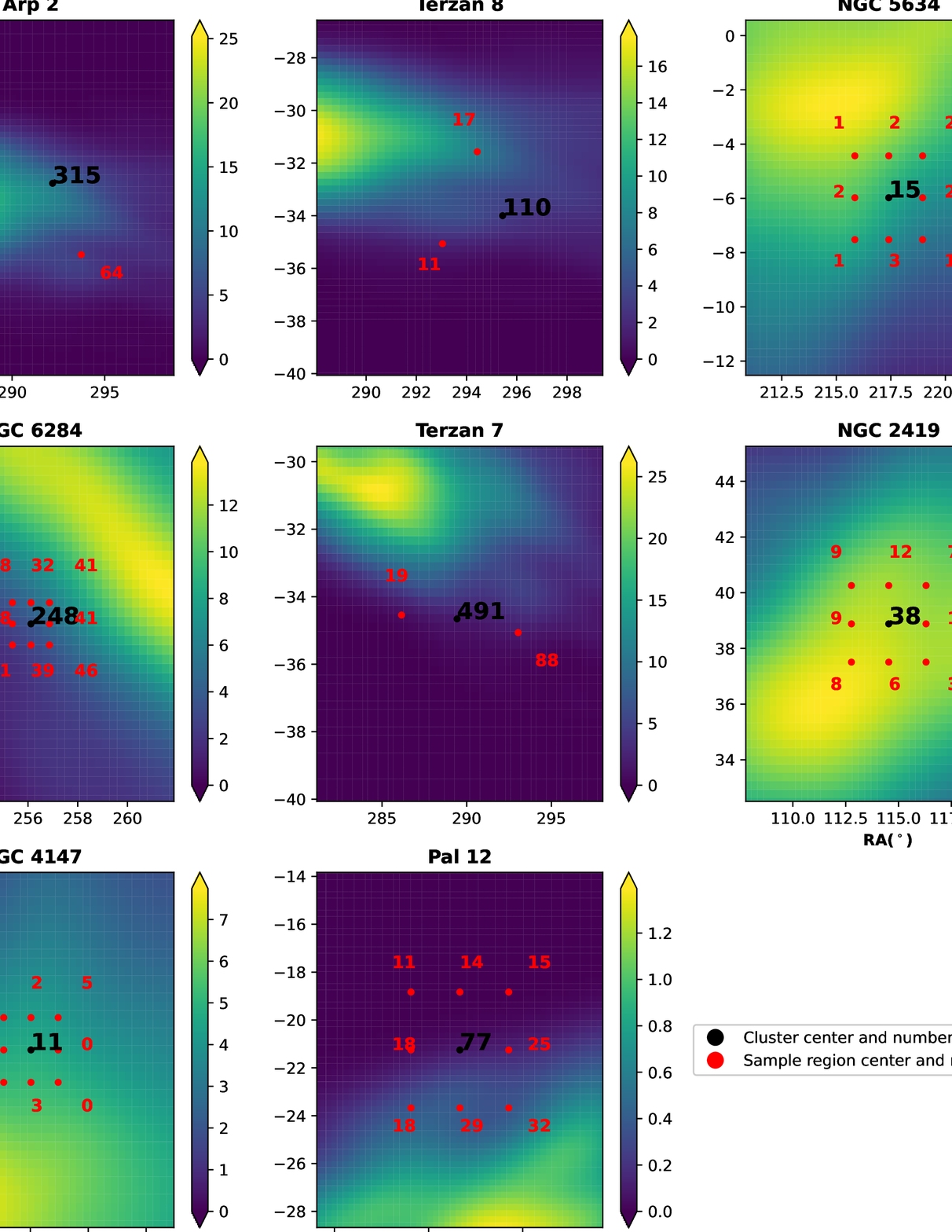}
\caption{Density map (stars per square degree) of Sgr dSph particles based on the LM10 model, along with the clusters (black) and the number of contaminants found around each cluster, and the scaled number of stars in each sample region (red), along with the corresponding number of extra-tidal candidates (black number and white dots).}
\label{lm}
\end{center}
\end{figure*}

To get a full picture of the spatial distribution and orientation of the extra-tidal candidates, we overplotted their distribution over the spatial distribution of the LM10 model particles. Figure~\ref{lm} displays the location of the centre of each cluster and comparison fields, along with the scaled number of field stars\footnote{Scaled number of field stars = (number of selected stars in the region) $\times$ (total number of stars around the cluster) $\times$ (total number of stars in the region)$^{-1}$}. In Figure~\ref{lm}, each cluster region is plotted including the particles from the model of LM10 that were likely stripped during the same perigalactic passage as each cluster (same Pcol as in Table 1 of B20). We compare the spatial distribution of the extra-tidal candidates with that of the LM10 model particle in that region to get a better idea if they are both governed by the same forces or not.
\\

\section{Results and Discussion}

extra-tidal star candidates are detected for nine GCs. We find 315, 110, 15, 248, 491, 38, 11, 233, and 77 stars for Arp 2, Terzan 8, NGC 5634, NGC 6284, Terzan 7, NGC 2419, NGC 4147, M 54, and Pal 12, respectively. Most of these stars lie on the RGB of the respective cluster CMD, except for stars belonging to NGC 4147 and Pal 12. For these two clusters, most of the stars lie on the main sequence of the clusters' CMDs. {\it Gaia} photometry is good enough to get the CMDs of these clusters up to the main sequence turn-off, except for NGC 2419, which is the most distant cluster in our sample. HB extra-tidal candidates are detected for four of them, namely, Arp 2 (20 stars), Terzan 8 (seven stars), NGC 6284 (three stars), and M 54 (22 stars). Two of these HB stars are classified as RR Lyrae variable stars in the {\it Gaia} DR2 catalogue \citep{gaiadr2rrl}; one star at 3.1 times $r_t$ from M 54\footnote{Gaia ID 6761169716763303424.}, and one star at 4.3 times $r_t$ from Terzan 8\footnote{Gaia ID 6741566416549068288.}. Given their position in the CMD, they are high-confidence extra-tidal stars from each cluster.
\\

\subsection{Level of contamination in the extra-tidal star candidates}

We estimated the level of contamination in the number of extra-tidal candidates detected for each cluster by selecting sample regions in different directions around them and applying the same criteria we used to select the extra-tidal candidates. The distance between the centre of the cluster and the centre of the sample region was at least 10$r_t$. This approach helps not only to detect the number of stars that may randomly fall under our selection criteria, but also to detect the presence of any extended tidal tail features around the cluster if there is an overdensity of stars in a preferred direction that could be connected to the overdensities recovered in the outskirts of the clusters. 
\\

It can be seen from Figure~\ref{lm} that the stars that pass our criteria to be selected as candidate extra-tidal stars are significantly more numerous (per unit area) in the surroundings of the cluster that in the control fields. This indicates that the selected extra-tidal candidates may have a very low level of contamination from MW foreground or background stars. This also indicates that the clusters do not seem to have any extended tidal-tails and the detected extra-tidal candidates seem to have a low level of contamination. This result could be due to the PM of the field stars in the sample regions (which are more than 10$r_t$  from the cluster centre), being significantly different from the PM of the cluster. However, some non-negligible contamination from the Sgr stream population could be present in the location of the extra-tidal stars.

\subsection{Radial velocities}

We expect the RV values of the stars belonging to a cluster to be concentrated around the mean RV of the cluster with small dispersion. Hence, a literature search was performed to get the RVs for the extra-tidal candidates to confirm their association with the corresponding clusters. Out of the 1538 extra-tidal candidates, we found RVs for eight stars from the extra-tidal star candidates around four of the studied clusters. In particular, one star with RV was found around NGC 2419 \citep[from][]{Ibata11}, three stars around NGC 4147 \citep[from][]{Kimmig15}, three stars around M 54 \citep[from][]{apogee}, and one star around Terzan 7 \citep[from][]{Frinchaboy12} with measured RV values. 

\begin{figure*}
\begin{center}
\includegraphics[width=\textwidth, height=\textheight, keepaspectratio]{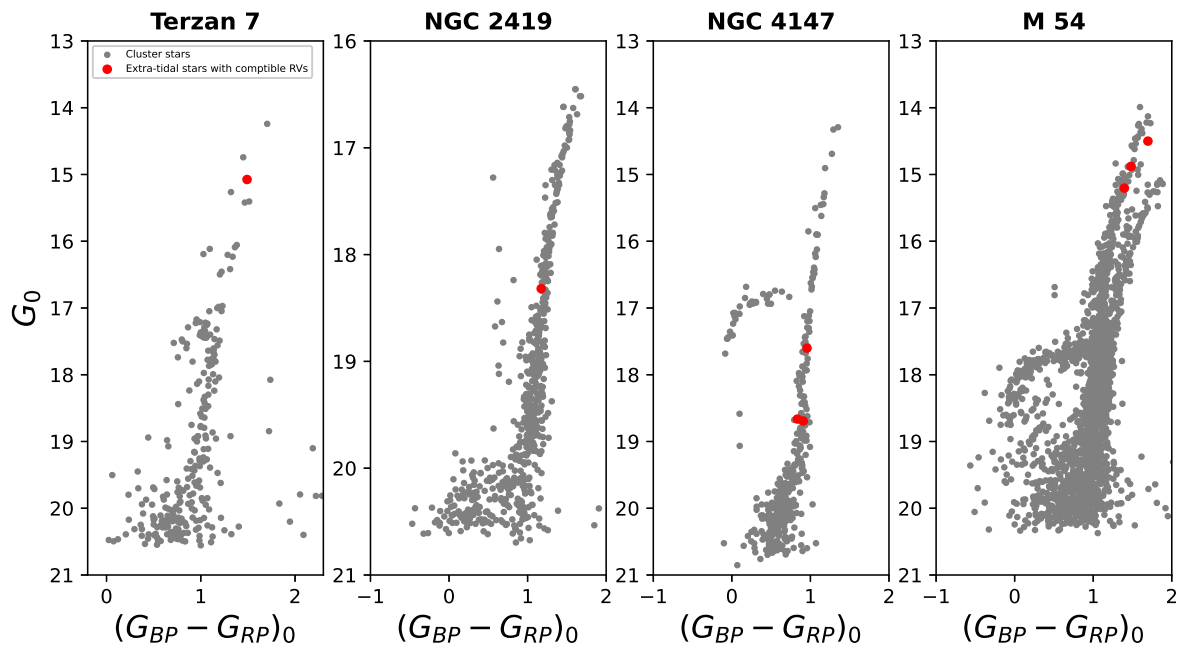}
\caption{Position of extra-tidal stars with compatible RVs on the respective {\it Gaia} EDR3 cluster CMDs.}
\label{hr_rv}
\end{center}
\end{figure*}

\begin{figure*}
\begin{center}
\includegraphics[width=\textwidth, height=\textheight, keepaspectratio]{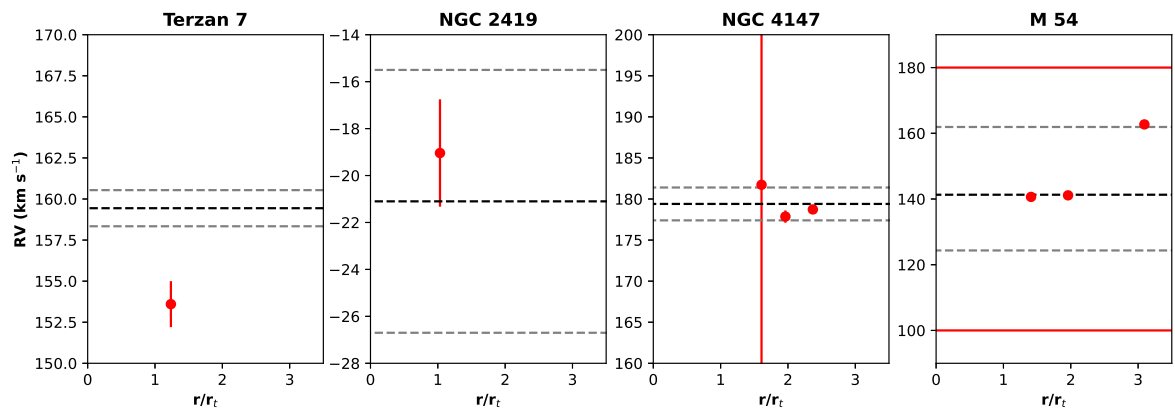}
\caption{Measured RVs of the stars with respect to their distances from cluster centres over the $r_t$ of each cluster. Red dots are the stars having RV measurements in the literature. The grey line is used for the mean RVs of the clusters, while the black lines represent the RV dispersion. Red lines in the M 54 panel are the limits on the RVs adopted by \citet{Bellazzini08}. The closest star to NGC 4147 has a larger RV error of $\sim$ 80 km s$^{-1}$, extending beyond the limits of the plot.}
\label{pos_rv}
\end{center}
\end{figure*}

We considered a star as being compatible with the cluster population if its RV was within the RV dispersion of the cluster. The adopted mean RV and its dispersion value are those reported by \citet{Baumgardt19}. The mean RV for NGC 2419 is --21.1 $\pm$0.3 km s$^{-1}$ ($\sigma_{\rm RV}$ = 5.6 km s$^{-1}$), and so the star with RV equal to --19.04$\pm$2.29 km s$^{-1}$ was considered compatible with this cluster. The mean RV for NGC 4147 is 179.4$\pm$0.3 km s$^{-1}$ ($\sigma_{\rm RV}$ = 2.0 km s$^{-1}$), and all three of the stars have RVs within the RV dispersion of the cluster, although one star has a large RV error of $\sim$80 km s$^{-1}$.  In order to select the compatible stars for M 54, we adopted the same limits as those adopted by \citet{Bellazzini08}, and stars with RVs between 100 km s$^{-1}$ and 180 km s$^{-1}$ were considered likely compatible with the cluster. Terzan 7 has a mean RV of 159.4$\pm$0.4 km s$^{-1}$ ($\sigma_{\rm RV}$ = 1.1 km \texttt{s$^{-1}$}), and the extra-tidal candidate star has an RV value of 153.6$\pm$1.4 km s$^{-1}$, having no compatible RV with the cluster motion for only a few km s$^{-1}$. Hence, based on this comparison, all but one of these eight stars have compatible RVs with their respective clusters. The positions of these stars on the respective {\it Gaia} EDR3 CMDs are shown in Figure~\ref{hr_rv}. Figure~\ref{pos_rv} shows the RVs of these eight stars with respect to their distances from the cluster centre. The grey line marks the mean RV of each cluster and the black lines represent the RV dispersion of the cluster. Red lines in the M 54 panel are the limits on the RV adopted by \citet{Bellazzini08}. The spatial distribution of the selected stars is shown as white dots in Figure~\ref{map}.

In the case of NGC 2419, the only extra-tidal star with a compatible RV is close to the $r_t$ of the cluster. Unfortunately, given the large distance of NGC~2419, there are very few studies in the literature that include RV measurements, and therefore no firm conclusions can be derived from this search. Conversely, for M 54 there are stars up to 4 $r_t$ from the cluster centre, all of them having consistent RV. However, as the cluster is located inside the main body of Sgr dSph, it is difficult to separate the stars from the cluster itself from the stars that are part of the galaxy. Chemical abundances for these stars, which may become available in the future, could be the key to unambiguously separating both populations. In the case of NGC 4147, all the stars with compatible RVs are up to 15 arcmin from the cluster centre. This is already noticed in the work of \citet{Hanke20}, in which several extra-tidal stars are found, based on a chemodynamical search over the whole sky. However, inside 5 $r_t$, \citet{Hanke20} only recover one star with a compatible RV. We have two additional stars from the sample of \citet{Kimmig15}. Having RVs in the literature compatible with the motion of the clusters for seven out of the eight extra-tidal star candidates with previous measurements confirms the fact that our candidate stars are most likely extra-cluster stars. RV measurements for all the candidates presented in this work are needed to confirm their individual nature.

\subsection{Possible reasons for the presence of extra-tidal stars around the clusters}

The list of extra-tidal candidates published in this work is not complete because, in an attempt to keep the sample as free as possible from false positives, we may have lost some genuine candidates. Once the RV measurements are available for all the candidates, it might be possible to confirm  the overdensities detected in this work, and a better understanding of their origin could be assessed. Nonetheless, given the preliminary results obtained in this work, in particular those for the orientation of the overdensities of these candidates with respect the Sgr main body position, the Galactic centre, and the PM of the clusters, can be used for a cluster-by-cluster analysis. This could be further used to elucidate the reasons behind the presence of the detected extra-tidal star candidates.
\\

\citet{Gnedin97} studied the effect of dynamical friction, relaxation, bulge, and disk shocks on GCs. Based on the destruction rates they provided, Arp 2, NGC 6284, M 54, and Pal 12 would be destroyed by bulge and disk shocks faster than by the other processes, while the rest of the clusters are equally affected by both relaxation and shocks. We analysed the destruction of these clusters based on the spatial distribution of the extra-tidal candidates around them. 
\\

{\bf Arp 2:} Out of the 315 extra-tidal candidates, 271 are outside the cluster's $r_J$. The density map, based on the extra-tidal candidates detected in this analysis, shows an overdensity contour towards the south-western direction. The overdensity is aligned with the PM in the leading direction. The morphology of extra-tidal candidates pointing towards and away from the cluster motion is typically due to tidal disruptions \citep{leon00, kuzma15}. However no overdensity in the opposite direction is seen. The core of Sgr dSph also lies on the western side (see Figure~\ref{lm}) of the cluster. Hence, an explanation for these extra-tidal stars could be that these stars are being stripped apart due to the tidal forces acting on the cluster, and the gravitational force from the core of the Sgr is increasing its rate of destruction. It is worth mentioning that the level of possible contaminants among the extra-tidal stars (e.g. field stars with compatible PM and CMD positions) is very small, based on the scaled number of field stars in two adjacent, iso-density fields (see Section 4.2). Therefore, this large number of extra-tidal stars could indicate that the cluster is suffering more disruption than the other clusters in the main body and/or the adopted $r_t$ is underestimated. \citet{Salinas12} also find an excess of blue stragglers beyond the $r_t$ of the cluster, and conclude that the cluster has already relaxed and is no longer under the process of two-body relaxation.
\\

{\bf Terzan 8:} 80 out of 110 detected extra-tidal stars are found outside the $r_J$ of the cluster. The scaled number of field stars is negligible compared to the number of detected extra-tidal stars, which indicates a low level of contamination. The extra-tidal stars seem to be evenly distributed around the cluster in the region outside its $r_J$. This is an indication that the cluster may be most affected by the internal relaxation \citep{kuzma16, kuzma17}. However, a small overdensity aligned towards the leading direction of the PM of the cluster is seen inside its $r_J$. This overdensity of potential escapers is very similar to the one seen near Arp 2. It seems that the cluster is equally affected by internal relaxation and tidal forces, and being near to the core of the Sgr increases its destruction rate. \citet{Carretta14} find a few stars with compatible RVs at distances from the cluster centre greater than 4 arcmin, up to 5.8 arcmin, considering all of them part of the cluster population. \citet{Salinas12} find eight HBs lying outside the $r_t$ of the cluster, which may point towards a possible disruption of the cluster. Our findings support this scenario as we find extra-tidal candidates up to 19.8 arcmin.
\\

{\bf NGC 5634:} Five of the extra-tidal candidates detected for the cluster lie inside the $r_J$ and the other ten lie outside. Although the number of extra-tidal stars around the cluster is small, the scaled number of field stars around this cluster is even smaller, indicating a low level of contamination among the extra-tidal stars. Based on the density map of the cluster, it seems to have two overdensities (within 5$r_t$), aligned with the direction of the Sgr stream particles from the LM10 model (see Figure~\ref{lm}), extending in the south-eastern and north-western directions. This could mean that the cluster is associated with Sgr dSph (also concluded by B20) as the extra-tidal candidates in the vicinity seem to be affected by the same gravitational field that is affecting the Sgr stream. \citet{Julio14} study this cluster (purely based on photometric data) and conclude that no extra-tidal features are seen around the cluster at the same distance as that of the cluster, but a similar population is seen in the background which is found to be two times more distant than the cluster itself. 
\\

{\bf NGC 6284:} Based on B20, this cluster may not be associated with Sgr dSph. Nevertheless, we include it here for the sake of completeness of their published list of clusters. Its adopted $r_J$ is a bit smaller than its $r_t$, hence all the extra-tidal candidates should be unbound from the cluster. We again find a small scaled number of field stars around the cluster, indicating a low level of contamination. The number of extra-tidal stars is the largest among the clusters outside the main body of Sgr, reaching up to 248 stars. The density map of the extra-tidal candidates reveals that their overdensity contours point in the north-south direction. This does not seem to be aligned with the Sgr stream in that direction (based on the particles from the LM10 model, see Figure~\ref{lm}). Based on B20 analysis, this cluster is not associated with Sgr dSph, which is also evident from the non-alignment of the extra-tidal stars with the Sgr stream particles. The morphology of the overdensity seems to resemble the typical S-shape of tidally stripped stars in the inner contour. Based on the destruction rates estimated in \citet{Gnedin97}, this cluster is more prone to bulge and disk shocks, as its orbit crosses the Galactic plane \citep{Baumgardt19}. If the overdensity were confirmed, this would be the first evidence of tidal tails emerging from the cluster, as far as we are aware.
\\

{\bf Terzan 7:} The adopted $r_t$ of the cluster is slightly smaller than its $r_J$, and therefore out of the 491 extra-tidal candidates, 484 are unbound from the cluster. The scaled number of field stars around the cluster again indicates a low level of contamination. The density map in Figure~\ref{map} shows that the distribution of the extra-tidal candidates is mainly concentrated in the north-western semicircle around the cluster with an overdensity lying on the western side, between the PM direction and the direction towards the core. This is consistent with the distribution of the LM10 particles in the region (see Figure ~\ref{lm}). This may indicate that the extra-tidal region of the cluster may be mainly affected by the gravitational force of the Sgr dSph core. We find RV measurements for one of the extra-tidal candidates from \citet{Frinchaboy12}, having an almost compatible RV with the cluster motion, being less than 5 km s$^{-1}$ offset from the mean RV $\pm$ RV dispersion of the cluster. We therefore consider this star as likely being compatible with its cluster. Although the star is outside the overdensities found in Figure 3, it is still located on the north-western side of the cluster just outside its $r_J$.
\\

{\bf NGC 2419:} 12 stars out of the 38 extra-tidal candidates are outside the $r_J$ of the cluster. Again, the scaled number of field stars around the cluster indicates that the extra-tidal candidates have a low contamination level. The density map of the region shows that the extra-tidal candidates outside the $r_J$ are almost evenly distributed, without any preferred direction, forming an extended stellar envelope around the cluster \citep{kuzma16, kuzma17}. However, an overdensity of potential escapers is seen towards the western side of the cluster. This is again aligned with the direction of the Sgr stream, which indicates the association of the cluster with Sgr. The overdensity in the north-western direction, in the vicinity of the cluster, is also detected by \citet{Jordi10}, which could indicate that the cluster is dissolving with time. Previous studies on the presence of extra-tidal stars around the cluster did not agree on a single result  \citep{Ripepi07, Bellazzini07}. We also find RVs corresponding to one of the extra-tidal candidates in the literature \citep{Ibata11} and it is compatible with the mean RV of the cluster, which is reassuring as this is the most distant cluster of the sample and, therefore, the {\it Gaia} data astrometry for those stars is less precise.
\\

{\bf NGC 4147:} Out of the 11 candidates, six lie outside the $r_J$ of the cluster and hence might be completely detached from it. The density map obtained for this cluster shows that the overdensity of extra-tidal candidates also lie inside the $r_J$. The scaled number of field stars around the cluster indicates a low level of contamination except in the north-western direction, which is also the direction of the overdensity. The cluster may have lost most of its loosely bound stars in the past and, presently, its stars are weakly affected by the gravitational forces affecting the Sgr stream. \citet{Jordi10} also find an overdensity around the cluster in north-south direction, but our analysis only reveals one overdensity on the northern side. We find RVs for three extra-tidal candidates in the literature \citep{Kimmig15} and all of them are compatible with the RV of the cluster. The distribution of these stars (see Figure~\ref{map}) is very similar to the overdensity contour in the density map. The overdensity seen in this map is in the direction opposite to its motion and may be due to disk and/or bulge shocks.
\\

{\bf M 54:} This cluster almost coincides with the core of Sgr dSph galaxy in 3D space. The density map of these extra-tidal candidates shows an overdensity towards the western side of the cluster. We have searched up to 5$r_t$ of the cluster and the $r_J$ is more than our search radius. Hence, although these stars are outside the $r_t$ of the cluster, they may not be out of its gravitational potential. The PM of the cluster is almost the same as the PM of the Sgr core and, therefore, we expect a lot of contamination among the selected extra-tidal stars around the cluster. However,  its unique position makes it almost impossible for us to estimate the level of contamination in the extra-tidal candidates based solely on the photometric and astrometric data. We find RV measurements for three candidates \citep{apogee} and all of them have RVs compatible with the cluster up to almost four $r_t$ from the cluster centre ($\sim$25 arcmin). This population of extra-tidal stars could extend even beyond the area studied in this work, although the identification of these stars depends on how different the PMs are from the cluster and the Sgr stream stars. Another approach to identify extra-tidal stars is based on peculiar chemical abundances. In fact, \citet{jose21} report a serendipitous discovery of a N-rich extra-tidal star at 45 arcmin (6 $r_t$) from the centre of M~54 with consistent chemical abundances as well as kinematics, including PM and RVs.
\\ 

{\bf Pal 12}: The adopted $r_t$ of the cluster is larger than its $r_J$ and, therefore, all the 77 extra-tidal star candidates are gravitationally unbound from it. The overdensities are present in the southern part of the search radius (see the density map in Figure~\ref{map}). This direction is coincident with the location of the Sgr stream stars (based on the LM10 particles). Our results seem to be in good agreement with the scenario in which the cluster is associated with Sgr dSph, as both the cluster stars and the stream are aligned. Based on \citet[][and B20]{Musella18} works, we a expect significant overdensity of Sgr stream stars in the same direction as the one detected in our analysis.  Nonetheless, the PM of the cluster can be easily separated from the stream stars (see Fig 4 in B20). Therefore, it is unlikely to have a significant number of Sgr stream stars to be included in the sample of extra-tidal star candidates recovered. In fact, from our analysis based on sample fields, a low contamination level is also indicated. In previous photometric studies, \citet{leon00} reported extended tidal tails around this cluster, and a larger $r_t$ based on the King profile. These results were later discarded by \citet{Musella18}, who find that the $r_t$ was overestimated due to an overdensity of Sgr stream stars, while no significant overdensity of extra-tidal Pal 12 stars was found. Our results agree with those of \citet{Musella18} as there are no prominent, significant tidal tails emerging from this low-mass cluster, although there could be some unbound stars due to its stripping from the Sgr galaxy in the last peri-centric passage.

\section{Conclusion}

We analysed the extra-tidal regions of nine GCs possibly associated with Sgr dSph. We selected the extra-tidal candidates based on their position in the RA-Dec plane, on the PMs of the clusters and the stars, and on the position of the stars on the de-reddened CMDs of the clusters. We estimated the mean PMs and PM dispersion of the clusters using a superposition of two Gaussian models (one for the field stars and the other for the cluster population). We found extra-tidal candidates in nine of the clusters and determined the contamination levels around eight of them (excluding M 54 due to its unique position). For the contamination analysis, we selected two sample regions around the central clusters and eight sample regions around the rest of the clusters lying in different directions. This approach also enabled us to identify any real (not composed of field stars) extended extra-tidal features in a preferred direction that could be present around them. However, none of the fields had a scaled number of contaminants comparable with the number of extra-tidal stars around the corresponding cluster.

Clusters that are closer to the core of Sgr dSph (namely Arp 2, Terzan 7, and Terzan 8) seem to be more affected by Sgr dSph gravitational potential than the farthest clusters (NGC 4147, NGC 2419 and NGC 6284). Based on the findings by B20, NGC 5634 (likely member) and Pal 12 (confirmed member) are associated with Sgr dSph, whereas, NGC 6284  is most probably not associated with Sgr dSph, and the results based on the spatial distribution of the detected extra-tidal candidates are in agreement.

We do not find any extended (outside 5$r_t$) extra-tidal features around any of the clusters. The reason for this could be that the stars that were lost in the past are no longer traceable based on the PMs and CMDs of the clusters, or the density of the stars is too low to be recovered based on RGB stars. They could have joined the Sgr stream and thus are no longer moving with the PM that might be related to the cluster. However, chemical analysis of stars in the regions surrounding these clusters might enable the discovery of such objects due to their similarity with the composition of the parent cluster.

To conclude, it is worth mentioning that all the candidates found in this work need to be confirmed by a star-by-star analysis of their RVs. Nonetheless, these are the most probable candidates based on the Gaia astrometry and photometry data available to date. Based on the adopted cuts, this list may not be complete as we may have lost a few of the real candidates at the expense of keeping the data free from false positives as much as possible. However, the membership of these stars could be confirmed once their RVs are available from future {\it Gaia} releases or spectroscopic follow-ups. Having the full 6D phase-space information for these candidates is crucial in order to really assess their origin and extension.

\begin{acknowledgements}

Authors would like to thank the referee for constructive and useful comments which improved the quality of the draft significantly. We would like to thank M. Bellazzini for promptly providing the Sgr dSph population selection from B20. R.~K and D.~M are very grateful for the hospitality of the Vatican Observatory, where this collaboration was started. D.M. gratefully acknowledges support by the ANID BASAL projects ACE210002 and FB210003, and by Fondecyt Project No. 1220724. This research has made use of the NASA/IPAC Infrared Science Archive, which is funded by the National Aeronautics and Space Administration and operated by the California Institute of Technology. HPS acknowledges grant 03(1428)/18/EMR-II from Council of Scientific and Industrial Research (CSIR), India. J.G.F-T gratefully acknowledges the grant support provided by Proyecto Fondecyt Iniciaci\'on No. 11220340, and also from ANID Concurso de Fomento a la Vinculaci\'on Internacional para Instituciones de Investigaci\'on Regionales (Modalidad corta duraci\'on) Proyecto No. FOVI210020, and from the Joint Committee ESO-Government of Chile 2021 (ORP 023/2021). R.K. would also like to thank Riyal Rana, her daughter, who was born while the paper was still in the process, for giving her so much happiness.\\

\end{acknowledgements}





\bibliographystyle{aa}
\bibliography{sgr} 

\begin{thebibliography}{112}
\expandafter\ifx\csname natexlab\endcsname\relax\def\natexlab#1{#1}\fi

\bibitem[{{Akaike}(1974)}]{Akaike74}
{Akaike}, H. 1974, IEEE Transactions on Automatic Control, 19, 716

\bibitem[{{Antoja} {et~al.}(2020){Antoja}, {Ramos}, {Mateu}, {Helmi}, {Anders},
  {Jordi}, \& {Carballo-Bello}}]{antoja20}
{Antoja}, T., {Ramos}, P., {Mateu}, C., {et~al.} 2020, A\&A, 635, L3

\bibitem[{{Baumgardt} {et~al.}(2019){Baumgardt}, {Hilker}, {Sollima}, \&
  {Bellini}}]{Baumgardt19}
{Baumgardt}, H., {Hilker}, M., {Sollima}, A., \& {Bellini}, A. 2019, MNRAS,
  482, 5138

\bibitem[{{Baumgardt} {et~al.}(2010){Baumgardt}, {Parmentier}, {Gieles}, \&
  {Vesperini}}]{Baumgardt10}
{Baumgardt}, H., {Parmentier}, G., {Gieles}, M., \& {Vesperini}, E. 2010,
  MNRAS, 401, 1832

\bibitem[{{Bellazzini}(2007)}]{Bellazzini07}
{Bellazzini}, M. 2007, A\&A, 473, 171

\bibitem[{{Bellazzini} {et~al.}(2003){Bellazzini}, {Ferraro}, \&
  {Ibata}}]{Bellazzini03}
{Bellazzini}, M., {Ferraro}, F.~R., \& {Ibata}, R. 2003, AJ, 125, 188

\bibitem[{{Bellazzini} {et~al.}(2020){Bellazzini}, {Ibata}, {Malhan}, {Martin},
  {Famaey}, \& {Thomas}}]{Bellazzini20}
{Bellazzini}, M., {Ibata}, R., {Malhan}, K., {et~al.} 2020, A\&A, 636, A107

\bibitem[{{Bellazzini} {et~al.}(2008){Bellazzini}, {Ibata}, {Chapman},
  {Mackey}, {Monaco}, {Irwin}, {Martin}, {Lewis}, \&
  {Dalessandro}}]{Bellazzini08}
{Bellazzini}, M., {Ibata}, R.~A., {Chapman}, S.~C., {et~al.} 2008, AJ, 136,
  1147

\bibitem[{{Belokurov} {et~al.}(2014){Belokurov}, {Koposov}, {Evans},
  {Pe{\~n}arrubia}, {Irwin}, {Smith}, {Lewis}, {Gieles}, {Wilkinson},
  {Gilmore}, {Olszewski}, \& {Niederste-Ostholt}}]{Belokurov14}
{Belokurov}, V., {Koposov}, S.~E., {Evans}, N.~W., {et~al.} 2014, MNRAS, 437,
  116

\bibitem[{{Belokurov} {et~al.}(2006){Belokurov}, {Zucker}, {Evans}, {Gilmore},
  {Vidrih}, {Bramich}, {Newberg}, {Wyse}, {Irwin}, {Fellhauer}, {Hewett},
  {Walton}, {Wilkinson}, {Cole}, {Yanny}, {Rockosi}, {Beers}, {Bell},
  {Brinkmann}, {Ivezi{\'c}}, \& {Lupton}}]{Belokurov06}
{Belokurov}, V., {Zucker}, D.~B., {Evans}, N.~W., {et~al.} 2006, ApJL, 642,
  L137

\bibitem[{{Bonifacio} {et~al.}(2004){Bonifacio}, {Sbordone}, {Marconi},
  {Pasquini}, \& {Hill}}]{Bonifacio04}
{Bonifacio}, P., {Sbordone}, L., {Marconi}, G., {Pasquini}, L., \& {Hill}, V.
  2004, A\&A, 414, 503

\bibitem[{Bressan {et~al.}(2012)Bressan, Marigo, Girardi, Salasnich, Dal~Cero,
  Rubele, \& Nanni}]{Bressan12}
Bressan, A., Marigo, P., Girardi, L., {et~al.} 2012, MNRAS, 427, 127

\bibitem[{Brown {et~al.}(1999)Brown, Wallerstein, \& Gonzalez}]{Brown1999}
Brown, J.~A., Wallerstein, G., \& Gonzalez, G. 1999, The Astronomical Journal,
  118, 1245

\bibitem[{{Brown} {et~al.}(1995){Brown}, {Wallerstein}, \& {Zucker}}]{brown95}
{Brown}, J.~A., {Wallerstein}, G., \& {Zucker}, D. 1995, in American
  Astronomical Society Meeting Abstracts, Vol. 187, American Astronomical
  Society Meeting Abstracts, 82.04

\bibitem[{{Bullock} \& {Boylan-Kolchin}(2017)}]{Bullock17r}
{Bullock}, J.~S. \& {Boylan-Kolchin}, M. 2017, ARAA, 55, 343

\bibitem[{{Carballo-Bello} {et~al.}(2017){Carballo-Bello}, {Corral-Santana},
  {Mart{\'\i}nez-Delgado}, {Sollima}, {Mu{\~n}oz}, {C{\^o}t{\'e}}, {Duffau},
  {Catelan}, \& {Grebel}}]{Julio17}
{Carballo-Bello}, J.~A., {Corral-Santana}, J.~M., {Mart{\'\i}nez-Delgado}, D.,
  {et~al.} 2017, MNRAS, 467, L91

\bibitem[{Carballo-Bello {et~al.}(2011)Carballo-Bello, Gieles, Sollima,
  Koposov, Martínez-Delgado, \& Peñarrubia}]{julio11}
Carballo-Bello, J.~A., Gieles, M., Sollima, A., {et~al.} 2011, MNRAS, 419, 14

\bibitem[{Carballo-Bello {et~al.}(2014)Carballo-Bello, Sollima,
  Martínez-Delgado, Pila-Díez, Leaman, Fliri, Muñoz, \&
  Corral-Santana}]{Julio14}
Carballo-Bello, J.~A., Sollima, A., Martínez-Delgado, D., {et~al.} 2014,
  MNRAS, 445, 2971

\bibitem[{{Carraro} {et~al.}(2007){Carraro}, {Zinn}, \& {Moni
  Bidin}}]{Carraro07}
{Carraro}, G., {Zinn}, R., \& {Moni Bidin}, C. 2007, A\&A, 466, 181

\bibitem[{{Carretta} {et~al.}(2009){Carretta}, {Bragaglia}, {Gratton},
  {D'Orazi}, \& {Lucatello}}]{Carretta09}
{Carretta}, E., {Bragaglia}, A., {Gratton}, R., {D'Orazi}, V., \& {Lucatello},
  S. 2009, A\&A, 508, 695

\bibitem[{{Carretta} {et~al.}(2014){Carretta}, {Bragaglia}, {Gratton},
  {D'Orazi}, {Lucatello}, \& {Sollima}}]{Carretta14}
{Carretta}, E., {Bragaglia}, A., {Gratton}, R.~G., {et~al.} 2014, A\&A, 561,
  A87

\bibitem[{{Carretta} {et~al.}(2017){Carretta}, {Bragaglia}, {Lucatello},
  {D'Orazi}, {Gratton}, {Donati}, {Sollima}, \& {Sneden}}]{Carretta17}
{Carretta}, E., {Bragaglia}, A., {Lucatello}, S., {et~al.} 2017, A\&A, 600,
  A118

\bibitem[{Claydon {et~al.}(2017)Claydon, Gieles, \& Zocchi}]{Claydon10}
Claydon, I., Gieles, M., \& Zocchi, A. 2017, MNARS, 466, 3937

\bibitem[{Clementini {et~al.}(2019)Clementini, Ripepi, Molinaro, Garofalo,
  Muraveva, Rimoldini, Guy, Jevardat~de Fombelle, Nienartowicz, Marchal,
  Audard, Holl, Leccia, Marconi, Musella, Mowlavi, Lecoeur-Taibi, Eyer,
  De~Ridder, Regibo, Sarro, Szabados, Evans, \& Riello}]{gaiadr2rrl}
Clementini, G., Ripepi, V., Molinaro, R., {et~al.} 2019, A\&A, 622, A60

\bibitem[{{Cohen}(2004)}]{cohen04}
{Cohen}, J.~G. 2004, AJ, 127, 1545

\bibitem[{{Da Costa} \& {Armandroff}(1995)}]{Costa95}
{Da Costa}, G.~S. \& {Armandroff}, T.~E. 1995, AJ, 109, 2533

\bibitem[{de Boer {et~al.}(2019)de Boer, Gieles, Balbinot, Hénault-Brunet,
  Sollima, Watkins, \& Claydon}]{Boer19}
de Boer, T. J.~L., Gieles, M., Balbinot, E., {et~al.} 2019, MNRAS, 485, 4906

\bibitem[{{Fern{\'a}ndez-Trincado} {et~al.}(2021){Fern{\'a}ndez-Trincado},
  {Beers}, {Minniti}, {Moni Bidin}, {Barbuy}, {Villanova}, {Geisler}, {Lane},
  {Roman-Lopes}, \& {Bizyaev}}]{jose21}
{Fern{\'a}ndez-Trincado}, J.~G., {Beers}, T.~C., {Minniti}, D., {et~al.} 2021,
  \aap, 648, A70

\bibitem[{{Forbes} \& {Bridges}(2010)}]{Forbes10}
{Forbes}, D.~A. \& {Bridges}, T. 2010, MNRAS, 404, 1203

\bibitem[{{Forbes} {et~al.}(2020){Forbes}, {Dullo}, {Gannon}, {Couch},
  {Iodice}, {Spavone}, {Cantiello}, \& {Schipani}}]{forbes20}
{Forbes}, D.~A., {Dullo}, B.~T., {Gannon}, J., {et~al.} 2020, MNRAS, 494, 5293

\bibitem[{{Frinchaboy} {et~al.}(2012){Frinchaboy}, {Majewski}, {Mu{\~n}oz},
  {Law}, {{\L}okas}, {Kunkel}, {Patterson}, \& {Johnston}}]{Frinchaboy12}
{Frinchaboy}, P.~M., {Majewski}, S.~R., {Mu{\~n}oz}, R.~R., {et~al.} 2012, ApJ,
  756, 74

\bibitem[{{Fukushige} \& {Heggie}(2000)}]{Fukushige2000}
{Fukushige}, T. \& {Heggie}, D.~C. 2000, MNRAS, 318, 753

\bibitem[{{Gaia Collaboration} {et~al.}(2021{\natexlab{a}}){Gaia
  Collaboration}, Antoja, McMillan, Kordopatis, Ramos, Helmi, Balbinot,
  Cantat-Gaudin, Chemin, Figueras, Jordi, Khanna, Romero-Gomez, \&
  Seabroke}]{antoja21}
{Gaia Collaboration}, Antoja, T., McMillan, P., {et~al.} 2021{\natexlab{a}},
  Gaia Early Data Release 3: The Galactic anticentre

\bibitem[{{Gaia Collaboration} {et~al.}(2021{\natexlab{b}}){Gaia
  Collaboration}, {Brown, A. G. A.}, {Vallenari, A.}, {Prusti, T.}, {de
  Bruijne, J. H. J.}, {Babusiaux, C.}, {Biermann, M.}, {Creevey, O. L.},
  {Evans, D. W.}, {Eyer, L.}, {Hutton, A.}, {Jansen, F.}, {Jordi, C.},
  {Klioner, S. A.}, {Lammers, U.}, {Lindegren, L.}, {Luri, X.}, {Mignard, F.},
  {Panem, C.}, {Pourbaix, D.}, {Randich, S.}, {Sartoretti, P.}, {Soubiran, C.},
  {Walton, N. A.}, {Arenou, F.}, {Bailer-Jones, C. A. L.}, {Bastian, U.},
  {Cropper, M.}, {Drimmel, R.}, {Katz, D.}, {Lattanzi, M. G.}, {van Leeuwen,
  F.}, {Bakker, J.}, {Cacciari, C.}, {Casta\~neda, J.}, {De Angeli, F.},
  {Ducourant, C.}, {Fabricius, C.}, {Fouesneau, M.}, {Fr\'emat, Y.}, {Guerra,
  R.}, {Guerrier, A.}, {Guiraud, J.}, {Jean-Antoine Piccolo, A.}, {Masana, E.},
  {Messineo, R.}, {Mowlavi, N.}, {Nicolas, C.}, {Nienartowicz, K.}, {Pailler,
  F.}, {Panuzzo, P.}, {Riclet, F.}, {Roux, W.}, {Seabroke, G. M.}, {Sordo, R.},
  {Tanga, P.}, {Th\'evenin, F.}, {Gracia-Abril, G.}, {Portell, J.}, {Teyssier,
  D.}, {Altmann, M.}, {Andrae, R.}, {Bellas-Velidis, I.}, {Benson, K.},
  {Berthier, J.}, {Blomme, R.}, {Brugaletta, E.}, {Burgess, P. W.}, {Busso,
  G.}, {Carry, B.}, {Cellino, A.}, {Cheek, N.}, {Clementini, G.}, {Damerdji,
  Y.}, {Davidson, M.}, {Delchambre, L.}, {Dell\'{}Oro, A.},
  {Fern\'andez-Hern\'andez, J.}, {Galluccio, L.}, {Garc\'{\i}a-Lario, P.},
  {Garcia-Reinaldos, M.}, {Gonz\'alez-N\'u\~nez, J.}, {Gosset, E.}, {Haigron,
  R.}, {Halbwachs, J.-L.}, {Hambly, N. C.}, {Harrison, D. L.}, {Hatzidimitriou,
  D.}, {Heiter, U.}, {Hern\'andez, J.}, {Hestroffer, D.}, {Hodgkin, S. T.},
  {Holl, B.}, {Jan\ss{}en, K.}, {Jevardat de Fombelle, G.}, {Jordan, S.},
  {Krone-Martins, A.}, {Lanzafame, A. C.}, {L\"offler, W.}, {Lorca, A.},
  {Manteiga, M.}, {Marchal, O.}, {Marrese, P. M.}, {Moitinho, A.}, {Mora, A.},
  {Muinonen, K.}, {Osborne, P.}, {Pancino, E.}, {Pauwels, T.}, {Petit, J.-M.},
  {Recio-Blanco, A.}, {Richards, P. J.}, {Riello, M.}, {Rimoldini, L.}, {Robin,
  A. C.}, {Roegiers, T.}, {Rybizki, J.}, {Sarro, L. M.}, {Siopis, C.}, {Smith,
  M.}, {Sozzetti, A.}, {Ulla, A.}, {Utrilla, E.}, {van Leeuwen, M.}, {van
  Reeven, W.}, {Abbas, U.}, {Abreu Aramburu, A.}, {Accart, S.}, {Aerts, C.},
  {Aguado, J. J.}, {Ajaj, M.}, {Altavilla, G.}, {\'Alvarez, M. A.}, {\'Alvarez
  Cid-Fuentes, J.}, {Alves, J.}, {Anderson, R. I.}, {Anglada Varela, E.},
  {Antoja, T.}, {Audard, M.}, {Baines, D.}, {Baker, S. G.},
  {Balaguer-N\'u\~nez, L.}, {Balbinot, E.}, {Balog, Z.}, {Barache, C.},
  {Barbato, D.}, {Barros, M.}, {Barstow, M. A.}, {Bartolom\'e, S.}, {Bassilana,
  J.-L.}, {Bauchet, N.}, {Baudesson-Stella, A.}, {Becciani, U.}, {Bellazzini,
  M.}, {Bernet, M.}, {Bertone, S.}, {Bianchi, L.}, {Blanco-Cuaresma, S.},
  {Boch, T.}, {Bombrun, A.}, {Bossini, D.}, {Bouquillon, S.}, {Bragaglia, A.},
  {Bramante, L.}, {Breedt, E.}, {Bressan, A.}, {Brouillet, N.}, {Bucciarelli,
  B.}, {Burlacu, A.}, {Busonero, D.}, {Butkevich, A. G.}, {Buzzi, R.}, {Caffau,
  E.}, {Cancelliere, R.}, {C\'anovas, H.}, {Cantat-Gaudin, T.}, {Carballo, R.},
  {Carlucci, T.}, {Carnerero, M. I}, {Carrasco, J. M.}, {Casamiquela, L.},
  {Castellani, M.}, {Castro-Ginard, A.}, {Castro Sampol, P.}, {Chaoul, L.},
  {Charlot, P.}, {Chemin, L.}, {Chiavassa, A.}, {Cioni, M.-R. L.}, {Comoretto,
  G.}, {Cooper, W. J.}, {Cornez, T.}, {Cowell, S.}, {Crifo, F.}, {Crosta, M.},
  {Crowley, C.}, {Dafonte, C.}, {Dapergolas, A.}, {David, M.}, {David, P.}, {de
  Laverny, P.}, {De Luise, F.}, {De March, R.}, {De Ridder, J.}, {de Souza,
  R.}, {de Teodoro, P.}, {de Torres, A.}, {del Peloso, E. F.}, {del Pozo, E.},
  {Delbo, M.}, {Delgado, A.}, {Delgado, H. E.}, {Delisle, J.-B.}, {Di Matteo,
  P.}, {Diakite, S.}, {Diener, C.}, {Distefano, E.}, {Dolding, C.}, {Eappachen,
  D.}, {Edvardsson, B.}, {Enke, H.}, {Esquej, P.}, {Fabre, C.}, {Fabrizio, M.},
  {Faigler, S.}, {Fedorets, G.}, {Fernique, P.}, {Fienga, A.}, {Figueras, F.},
  {Fouron, C.}, {Fragkoudi, F.}, {Fraile, E.}, {Franke, F.}, {Gai, M.},
  {Garabato, D.}, {Garcia-Gutierrez, A.}, {Garc\'{\i}a-Torres, M.}, {Garofalo,
  A.}, {Gavras, P.}, {Gerlach, E.}, {Geyer, R.}, {Giacobbe, P.}, {Gilmore, G.},
  {Girona, S.}, {Giuffrida, G.}, {Gomel, R.}, {Gomez, A.},
  {Gonzalez-Santamaria, I.}, {Gonz\'alez-Vidal, J. J.}, {Granvik, M.},
  {Guti\'errez-S\'anchez, R.}, {Guy, L. P.}, {Hauser, M.}, {Haywood, M.},
  {Helmi, A.}, {Hidalgo, S. L.}, {Hilger, T.}, {Hladczuk, N.}, {Hobbs, D.},
  {Holland, G.}, {Huckle, H. E.}, {Jasniewicz, G.}, {Jonker, P. G.}, {Juaristi
  Campillo, J.}, {Julbe, F.}, {Karbevska, L.}, {Kervella, P.}, {Khanna, S.},
  {Kochoska, A.}, {Kontizas, M.}, {Kordopatis, G.}, {Korn, A. J.},
  {Kostrzewa-Rutkowska, Z.}, {Kruszy\'{}nska, K.}, {Lambert, S.}, {Lanza, A.
  F.}, {Lasne, Y.}, {Le Campion, J.-F.}, {Le Fustec, Y.}, {Lebreton, Y.},
  {Lebzelter, T.}, {Leccia, S.}, {Leclerc, N.}, {Lecoeur-Taibi, I.}, {Liao,
  S.}, {Licata, E.}, {Lindstr\o{}m, E. P.}, {Lister, T. A.}, {Livanou, E.},
  {Lobel, A.}, {Madrero Pardo, P.}, {Managau, S.}, {Mann, R. G.}, {Marchant, J.
  M.}, {Marconi, M.}, {Marcos Santos, M. M. S.}, {Marinoni, S.}, {Marocco, F.},
  {Marshall, D. J.}, {Martin Polo, L.}, {Mart\'{\i}n-Fleitas, J. M.}, {Masip,
  A.}, {Massari, D.}, {Mastrobuono-Battisti, A.}, {Mazeh, T.}, {McMillan, P.
  J.}, {Messina, S.}, {Michalik, D.}, {Millar, N. R.}, {Mints, A.}, {Molina,
  D.}, {Molinaro, R.}, {Moln\'ar, L.}, {Montegriffo, P.}, {Mor, R.},
  {Morbidelli, R.}, {Morel, T.}, {Morris, D.}, {Mulone, A. F.}, {Munoz, D.},
  {Muraveva, T.}, {Murphy, C. P.}, {Musella, I.}, {Noval, L.}, {Ord\'enovic,
  C.}, {Orr\`u, G.}, {Osinde, J.}, {Pagani, C.}, {Pagano, I.}, {Palaversa, L.},
  {Palicio, P. A.}, {Panahi, A.}, {Pawlak, M.}, {Pe\~nalosa Esteller, X.},
  {Penttil\"a, A.}, {Piersimoni, A. M.}, {Pineau, F.-X.}, {Plachy, E.}, {Plum,
  G.}, {Poggio, E.}, {Poretti, E.}, {Poujoulet, E.}, {Prsa, A.}, {Pulone, L.},
  {Racero, E.}, {Ragaini, S.}, {Rainer, M.}, {Raiteri, C. M.}, {Rambaux, N.},
  {Ramos, P.}, {Ramos-Lerate, M.}, {Re Fiorentin, P.}, {Regibo, S.}, {Reyl\'e,
  C.}, {Ripepi, V.}, {Riva, A.}, {Rixon, G.}, {Robichon, N.}, {Robin, C.},
  {Roelens, M.}, {Rohrbasser, L.}, {Romero-G\'omez, M.}, {Rowell, N.}, {Royer,
  F.}, {Rybicki, K. A.}, {Sadowski, G.}, {Sagrist\`a Sell\'es, A.}, {Sahlmann,
  J.}, {Salgado, J.}, {Salguero, E.}, {Samaras, N.}, {Sanchez Gimenez, V.},
  {Sanna, N.}, {Santove\~na, R.}, {Sarasso, M.}, {Schultheis, M.}, {Sciacca,
  E.}, {Segol, M.}, {Segovia, J. C.}, {S\'egransan, D.}, {Semeux, D.}, {Shahaf,
  S.}, {Siddiqui, H. I.}, {Siebert, A.}, {Siltala, L.}, {Slezak, E.}, {Smart,
  R. L.}, {Solano, E.}, {Solitro, F.}, {Souami, D.}, {Souchay, J.}, {Spagna,
  A.}, {Spoto, F.}, {Steele, I. A.}, {Steidelm\"uller, H.}, {Stephenson, C.
  A.}, {S\"uveges, M.}, {Szabados, L.}, {Szegedi-Elek, E.}, {Taris, F.},
  {Tauran, G.}, {Taylor, M. B.}, {Teixeira, R.}, {Thuillot, W.}, {Tonello, N.},
  {Torra, F.}, {Torra, J.}, {Turon, C.}, {Unger, N.}, {Vaillant, M.}, {van
  Dillen, E.}, {Vanel, O.}, {Vecchiato, A.}, {Viala, Y.}, {Vicente, D.},
  {Voutsinas, S.}, {Weiler, M.}, {Wevers, T.}, {Wyrzykowski, L.}, {Yoldas, A.},
  {Yvard, P.}, {Zhao, H.}, {Zorec, J.}, {Zucker, S.}, {Zurbach, C.}, \&
  {Zwitter, T.}}]{gaiadr3}
{Gaia Collaboration}, {Brown, A. G. A.}, {Vallenari, A.}, {et~al.}
  2021{\natexlab{b}}, A\&A, 649, A1

\bibitem[{{Gaia Collaboration} {et~al.}(2016){Gaia Collaboration}, {Prusti},
  {de Bruijne}, {Brown}, {Vallenari}, {Babusiaux}, {Bailer-Jones}, {Bastian},
  {Biermann}, {Evans}, {Eyer}, {Jansen}, {Jordi}, {Klioner}, {Lammers},
  {Lindegren}, {Luri}, {Mignard}, {Milligan}, {Panem}, {Poinsignon},
  {Pourbaix}, {Randich}, {Sarri}, {Sartoretti}, {Siddiqui}, {Soubiran},
  {Valette}, {van Leeuwen}, {Walton}, {Aerts}, {Arenou}, {Cropper}, {Drimmel},
  {H{\o}g}, {Katz}, {Lattanzi}, {O'Mullane}, {Grebel}, {Holland}, {Huc},
  {Passot}, {Bramante}, {Cacciari}, {Casta{\~n}eda}, {Chaoul}, {Cheek}, {De
  Angeli}, {Fabricius}, {Guerra}, {Hern{\'a}ndez}, {Jean-Antoine-Piccolo},
  {Masana}, {Messineo}, {Mowlavi}, {Nienartowicz}, {Ord{\'o}{\~n}ez-Blanco},
  {Panuzzo}, {Portell}, {Richards}, {Riello}, {Seabroke}, {Tanga},
  {Th{\'e}venin}, {Torra}, {Els}, {Gracia-Abril}, {Comoretto},
  {Garcia-Reinaldos}, {Lock}, {Mercier}, {Altmann}, {Andrae}, {Astraatmadja},
  {Bellas-Velidis}, {Benson}, {Berthier}, {Blomme}, {Busso}, {Carry},
  {Cellino}, {Clementini}, {Cowell}, {Creevey}, {Cuypers}, {Davidson}, {De
  Ridder}, {de Torres}, {Delchambre}, {Dell'Oro}, {Ducourant}, {Fr{\'e}mat},
  {Garc{\'\i}a-Torres}, {Gosset}, {Halbwachs}, {Hambly}, {Harrison}, {Hauser},
  {Hestroffer}, {Hodgkin}, {Huckle}, {Hutton}, {Jasniewicz}, {Jordan},
  {Kontizas}, {Korn}, {Lanzafame}, {Manteiga}, {Moitinho}, {Muinonen},
  {Osinde}, {Pancino}, {Pauwels}, {Petit}, {Recio-Blanco}, {Robin}, {Sarro},
  {Siopis}, {Smith}, {Smith}, {Sozzetti}, {Thuillot}, {van Reeven}, {Viala},
  {Abbas}, {Abreu Aramburu}, {Accart}, {Aguado}, {Allan}, {Allasia},
  {Altavilla}, {{\'A}lvarez}, {Alves}, {Anderson}, {Andrei}, {Anglada Varela},
  {Antiche}, {Antoja}, {Ant{\'o}n}, {Arcay}, {Atzei}, {Ayache}, {Bach},
  {Baker}, {Balaguer-N{\'u}{\~n}ez}, {Barache}, {Barata}, {Barbier}, {Barblan},
  {Baroni}, {Barrado y Navascu{\'e}s}, {Barros}, {Barstow}, {Becciani},
  {Bellazzini}, {Bellei}, {Bello Garc{\'\i}a}, {Belokurov}, {Bendjoya},
  {Berihuete}, {Bianchi}, {Bienaym{\'e}}, {Billebaud}, {Blagorodnova},
  {Blanco-Cuaresma}, {Boch}, {Bombrun}, {Borrachero}, {Bouquillon}, {Bourda},
  {Bouy}, {Bragaglia}, {Breddels}, {Brouillet}, {Br{\"u}semeister},
  {Bucciarelli}, {Budnik}, {Burgess}, {Burgon}, {Burlacu}, {Busonero}, {Buzzi},
  {Caffau}, {Cambras}, {Campbell}, {Cancelliere}, {Cantat-Gaudin}, {Carlucci},
  {Carrasco}, {Castellani}, {Charlot}, {Charnas}, {Charvet}, {Chassat},
  {Chiavassa}, {Clotet}, {Cocozza}, {Collins}, {Collins}, {Costigan}, {Crifo},
  {Cross}, {Crosta}, {Crowley}, {Dafonte}, {Damerdji}, {Dapergolas}, {David},
  {David}, {De Cat}, {de Felice}, {de Laverny}, {De Luise}, {De March}, {de
  Martino}, {de Souza}, {Debosscher}, {del Pozo}, {Delbo}, {Delgado},
  {Delgado}, {di Marco}, {Di Matteo}, {Diakite}, {Distefano}, {Dolding}, {Dos
  Anjos}, {Drazinos}, {Dur{\'a}n}, {Dzigan}, {Ecale}, {Edvardsson}, {Enke},
  {Erdmann}, {Escolar}, {Espina}, {Evans}, {Eynard Bontemps}, {Fabre},
  {Fabrizio}, {Faigler}, {Falc{\~a}o}, {Farr{\`a}s Casas}, {Faye}, {Federici},
  {Fedorets}, {Fern{\'a}ndez-Hern{\'a}ndez}, {Fernique}, {Fienga}, {Figueras},
  {Filippi}, {Findeisen}, {Fonti}, {Fouesneau}, {Fraile}, {Fraser}, {Fuchs},
  {Furnell}, {Gai}, {Galleti}, {Galluccio}, {Garabato}, {Garc{\'\i}a-Sedano},
  {Gar{\'e}}, {Garofalo}, {Garralda}, {Gavras}, {Gerssen}, {Geyer}, {Gilmore},
  {Girona}, {Giuffrida}, {Gomes}, {Gonz{\'a}lez-Marcos},
  {Gonz{\'a}lez-N{\'u}{\~n}ez}, {Gonz{\'a}lez-Vidal}, {Granvik}, {Guerrier},
  {Guillout}, {Guiraud}, {G{\'u}rpide}, {Guti{\'e}rrez-S{\'a}nchez}, {Guy},
  {Haigron}, {Hatzidimitriou}, {Haywood}, {Heiter}, {Helmi}, {Hobbs},
  {Hofmann}, {Holl}, {Holland }, {Hunt}, {Hypki}, {Icardi}, {Irwin}, {Jevardat
  de Fombelle}, {Jofr{\'e}}, {Jonker}, {Jorissen}, {Julbe}, {Karampelas},
  {Kochoska}, {Kohley}, {Kolenberg}, {Kontizas}, {Koposov}, {Kordopatis},
  {Koubsky}, {Kowalczyk}, {Krone-Martins}, {Kudryashova}, {Kull}, {Bachchan},
  {Lacoste-Seris}, {Lanza}, {Lavigne}, {Le Poncin-Lafitte}, {Lebreton},
  {Lebzelter}, {Leccia}, {Leclerc}, {Lecoeur-Taibi}, {Lemaitre}, {Lenhardt},
  {Leroux}, {Liao}, {Licata}, {Lindstr{\o}m}, {Lister}, {Livanou}, {Lobel},
  {L{\"o}ffler}, {L{\'o}pez}, {Lopez-Lozano}, {Lorenz}, {Loureiro},
  {MacDonald}, {Magalh{\~a}es Fernandes}, {Managau}, {Mann}, {Mantelet},
  {Marchal}, {Marchant}, {Marconi}, {Marie}, {Marinoni}, {Marrese},
  {Marschalk{\'o}}, {Marshall}, {Mart{\'\i}n-Fleitas}, {Martino}, {Mary},
  {Matijevi{\v{c}}}, {Mazeh}, {McMillan}, {Messina}, {Mestre}, {Michalik},
  {Millar}, {Miranda}, {Molina}, {Molinaro}, {Molinaro}, {Moln{\'a}r},
  {Moniez}, {Montegriffo}, {Monteiro}, {Mor}, {Mora}, {Morbidelli}, {Morel},
  {Morgenthaler}, {Morley}, {Morris}, {Mulone}, {Muraveva}, {Musella},
  {Narbonne}, {Nelemans}, {Nicastro}, {Noval}, {Ord{\'e}novic},
  {Ordieres-Mer{\'e}}, {Osborne}, {Pagani}, {Pagano}, {Pailler}, {Palacin},
  {Palaversa}, {Parsons}, {Paulsen}, {Pecoraro}, {Pedrosa}, {Pentik{\"a}inen},
  {Pereira}, {Pichon}, {Piersimoni}, {Pineau}, {Plachy}, {Plum}, {Poujoulet},
  {Pr{\v{s}}a}, {Pulone}, {Ragaini}, {Rago}, {Rambaux}, {Ramos-Lerate},
  {Ranalli}, {Rauw}, {Read}, {Regibo}, {Renk}, {Reyl{\'e}}, {Ribeiro},
  {Rimoldini}, {Ripepi}, {Riva}, {Rixon}, {Roelens}, {Romero-G{\'o}mez},
  {Rowell}, {Royer}, {Rudolph}, {Ruiz-Dern}, {Sadowski}, {Sagrist{\`a}
  Sell{\'e}s}, {Sahlmann}, {Salgado}, {Salguero}, {Sarasso}, {Savietto},
  {Schnorhk}, {Schultheis}, {Sciacca}, {Segol}, {Segovia}, {Segransan},
  {Serpell}, {Shih}, {Smareglia}, {Smart}, {Smith}, {Solano}, {Solitro},
  {Sordo}, {Soria Nieto}, {Souchay}, {Spagna}, {Spoto}, {Stampa}, {Steele},
  {Steidelm{\"u}ller}, {Stephenson}, {Stoev}, {Suess}, {S{\"u}veges}, {Surdej},
  {Szabados}, {Szegedi-Elek}, {Tapiador}, {Taris}, {Tauran}, {Taylor},
  {Teixeira}, {Terrett}, {Tingley}, {Trager}, {Turon}, {Ulla}, {Utrilla},
  {Valentini}, {van Elteren}, {Van Hemelryck}, {van Leeuwen}, {Varadi},
  {Vecchiato}, {Veljanoski}, {Via}, {Vicente}, {Vogt}, {Voss}, {Votruba},
  {Voutsinas}, {Walmsley}, {Weiler}, {Weingrill}, {Werner}, {Wevers},
  {Whitehead}, {Wyrzykowski}, {Yoldas}, {{\v{Z}}erjal}, {Zucker}, {Zurbach},
  {Zwitter}, {Alecu}, {Allen}, {Allende Prieto}, {Amorim},
  {Anglada-Escud{\'e}}, {Arsenijevic}, {Azaz}, {Balm}, {Beck}, {Bernstein},
  {Bigot}, {Bijaoui}, {Blasco}, {Bonfigli}, {Bono}, {Boudreault}, {Bressan},
  {Brown}, {Brunet}, {Bunclark}, {Buonanno}, {Butkevich}, {Carret}, {Carrion},
  {Chemin}, {Ch{\'e}reau}, {Corcione}, {Darmigny}, {de Boer}, {de Teodoro}, {de
  Zeeuw}, {Delle Luche}, {Domingues}, {Dubath}, {Fodor}, {Fr{\'e}zouls},
  {Fries}, {Fustes}, {Fyfe}, {Gallardo}, {Gallegos}, {Gardiol}, {Gebran},
  {Gomboc}, {G{\'o}mez}, {Grux}, {Gueguen}, {Heyrovsky}, {Hoar}, {Iannicola},
  {Isasi Parache}, {Janotto}, {Joliet}, {Jonckheere}, {Keil}, {Kim},
  {Klagyivik}, {Klar}, {Knude}, {Kochukhov}, {Kolka}, {Kos}, {Kutka}, {Lainey},
  {LeBouquin}, {Liu}, {Loreggia}, {Makarov}, {Marseille}, {Martayan},
  {Martinez-Rubi}, {Massart}, {Meynadier}, {Mignot}, {Munari}, {Nguyen},
  {Nordlander}, {Ocvirk}, {O'Flaherty}, {Olias Sanz}, {Ortiz}, {Osorio},
  {Oszkiewicz}, {Ouzounis}, {Palmer}, {Park}, {Pasquato}, {Peltzer}, {Peralta},
  {P{\'e}turaud}, {Pieniluoma}, {Pigozzi}, {Poels}, {Prat}, {Prod'homme},
  {Raison}, {Rebordao}, {Risquez}, {Rocca-Volmerange}, {Rosen}, {Ruiz-Fuertes},
  {Russo}, {Sembay}, {Serraller Vizcaino}, {Short}, {Siebert}, {Silva},
  {Sinachopoulos}, {Slezak}, {Soffel}, {Sosnowska}, {Strai{\v{z}}ys}, {ter
  Linden}, {Terrell}, {Theil}, {Tiede}, {Troisi}, {Tsalmantza}, {Tur},
  {Vaccari}, {Vachier}, {Valles}, {Van Hamme}, {Veltz}, {Virtanen}, {Wallut},
  {Wichmann}, {Wilkinson}, {Ziaeepour}, \& {Zschocke}}]{gaiadr1}
{Gaia Collaboration}, {Prusti}, T., {de Bruijne}, J.~H.~J., {et~al.} 2016,
  A\&A, 595, A1

\bibitem[{{Garro} {et~al.}(2021){Garro}, {Minniti}, {G{\'o}mez}, \&
  {Alonso-Garc{\'\i}a}}]{Garro21}
{Garro}, E.~R., {Minniti}, D., {G{\'o}mez}, M., \& {Alonso-Garc{\'\i}a}, J.
  2021, arXiv e-prints, arXiv:2107.09987

\bibitem[{{Gnedin} \& {Ostriker}(1997)}]{Gnedin97}
{Gnedin}, O.~Y. \& {Ostriker}, J.~P. 1997, ApJ, 474, 223

\bibitem[{{Hamanowicz} {et~al.}(2016){Hamanowicz}, {Pietrukowicz}, {Udalski},
  {Mr{\'o}z}, {Soszy{\'n}ski}, {Szyma{\'n}ski}, {Skowron}, {Poleski},
  {Wyrzykowski}, {Koz{\l}owski}, {Pawlak}, \& {Ulaczyk}}]{Hamanowicz16}
{Hamanowicz}, A., {Pietrukowicz}, P., {Udalski}, A., {et~al.} 2016, \actaa, 66,
  197

\bibitem[{{Hanke} {et~al.}(2020){Hanke}, {Koch}, {Prudil}, {Grebel}, \&
  {Bastian}}]{Hanke20}
{Hanke}, M., {Koch}, A., {Prudil}, Z., {Grebel}, E.~K., \& {Bastian}, U. 2020,
  A\&A, 637, A98

\bibitem[{{Harris}(1996)}]{harris96}
{Harris}, W.~E. 1996, AJ, 112, 1487

\bibitem[{{Harris}(2010)}]{harris10}
{Harris}, W.~E. 2010, arXiv e-prints, arXiv:1012.3224

\bibitem[{{Hasselquist} {et~al.}(2019){Hasselquist}, {Carlin}, {Holtzman},
  {Shetrone}, {Hayes}, {Cunha}, {Smith}, {Beaton}, {Sobeck}, {Allende Prieto},
  {Majewski}, {Anguiano}, {Bizyaev}, {Garc{\'\i}a-Hern{\'a}ndez}, {Lane},
  {Pan}, {Nidever}, {Fern{\'a}ndez-Trincado}, {Wilson}, \&
  {Zamora}}]{Hasselquist19}
{Hasselquist}, S., {Carlin}, J.~L., {Holtzman}, J.~A., {et~al.} 2019, ApJ, 872,
  58

\bibitem[{{Hayes} {et~al.}(2020){Hayes}, {Majewski}, {Hasselquist}, {Anguiano},
  {Shetrone}, {Law}, {Schiavon}, {Cunha}, {Smith}, {Beaton}, {Price-Whelan},
  {Allende Prieto}, {Battaglia}, {Bizyaev}, {Brownstein}, {Cohen},
  {Frinchaboy}, {Garc{\'\i}a-Hern{\'a}ndez}, {Lacerna}, {Lane},
  {M{\'e}sz{\'a}ros}, {Bidin}, {M{\~{u}}noz}, {Nidever}, {Oravetz}, {Oravetz},
  {Pan}, {Roman-Lopes}, {Sobeck}, \& {Stringfellow}}]{Hayes20}
{Hayes}, C.~R., {Majewski}, S.~R., {Hasselquist}, S., {et~al.} 2020, ApJ, 889,
  63

\bibitem[{{Huxor} \& {Grebel}(2015)}]{Huxor15}
{Huxor}, A.~P. \& {Grebel}, E.~K. 2015, MNRAS, 453, 2653

\bibitem[{{Ibata} {et~al.}(2020){Ibata}, {Bellazzini}, {Thomas}, {Malhan},
  {Martin}, {Famaey}, \& {Siebert}}]{ibata20}
{Ibata}, R., {Bellazzini}, M., {Thomas}, G., {et~al.} 2020, ApJl, 891, L19

\bibitem[{{Ibata} {et~al.}(2011){Ibata}, {Sollima}, {Nipoti}, {Bellazzini},
  {Chapman}, \& {Dalessandro}}]{Ibata11}
{Ibata}, R., {Sollima}, A., {Nipoti}, C., {et~al.} 2011, ApJ, 738, 186

\bibitem[{{Ibata} {et~al.}(2019{\natexlab{a}}){Ibata}, {Bellazzini}, {Malhan},
  {Martin}, \& {Bianchini}}]{Ibata19a}
{Ibata}, R.~A., {Bellazzini}, M., {Malhan}, K., {Martin}, N., \& {Bianchini},
  P. 2019{\natexlab{a}}, Nature Astronomy, 3, 667

\bibitem[{{Ibata} {et~al.}(1994){Ibata}, {Gilmore}, \& {Irwin}}]{Ibata94}
{Ibata}, R.~A., {Gilmore}, G., \& {Irwin}, M.~J. 1994, Nature, 370, 194

\bibitem[{{Ibata} {et~al.}(1995){Ibata}, {Gilmore}, \& {Irwin}}]{ibata95}
{Ibata}, R.~A., {Gilmore}, G., \& {Irwin}, M.~J. 1995, MNRAS, 277, 781

\bibitem[{{Ibata} {et~al.}(2019{\natexlab{b}}){Ibata}, {Malhan}, \&
  {Martin}}]{Ibata19b}
{Ibata}, R.~A., {Malhan}, K., \& {Martin}, N.~F. 2019{\natexlab{b}}, ApJ, 872,
  152

\bibitem[{{Ibata} {et~al.}(2018){Ibata}, {Malhan}, {Martin}, \&
  {Starkenburg}}]{Ibata18}
{Ibata}, R.~A., {Malhan}, K., {Martin}, N.~F., \& {Starkenburg}, E. 2018, ApJ,
  865, 85

\bibitem[{{Johnston} {et~al.}(1995){Johnston}, {Spergel}, \&
  {Hernquist}}]{Johnston95}
{Johnston}, K.~V., {Spergel}, D.~N., \& {Hernquist}, L. 1995, ApJ, 451, 598

\bibitem[{{Jordi} \& {Grebel}(2010)}]{Jordi10}
{Jordi}, K. \& {Grebel}, E.~K. 2010, A\&A, 522, A71

\bibitem[{{Kimmig} {et~al.}(2015){Kimmig}, {Seth}, {Ivans}, {Strader},
  {Caldwell}, {Anderton}, \& {Gregersen}}]{Kimmig15}
{Kimmig}, B., {Seth}, A., {Ivans}, I.~I., {et~al.} 2015, AJ, 149, 53

\bibitem[{{King}(1962)}]{King62}
{King}, I. 1962, AJ, 67, 471

\bibitem[{{Koposov} {et~al.}(2012){Koposov}, {Belokurov}, {Evans}, {Gilmore},
  {Gieles}, {Irwin}, {Lewis}, {Niederste-Ostholt}, {Pe{\~n}arrubia}, {Smith},
  {Bizyaev}, {Malanushenko}, {Malanushenko}, {Schneider}, \&
  {Wyse}}]{Koposov12}
{Koposov}, S.~E., {Belokurov}, V., {Evans}, N.~W., {et~al.} 2012, ApJ, 750, 80

\bibitem[{{Kundu} {et~al.}(2019{\natexlab{a}}){Kundu},
  {Fern{\'a}ndez-Trincado}, {Minniti}, {Singh}, {Moreno}, {Reyl{\'e}}, {Robin},
  \& {Soto}}]{kundu19b}
{Kundu}, R., {Fern{\'a}ndez-Trincado}, J.~G., {Minniti}, D., {et~al.}
  2019{\natexlab{a}}, MNRAS, 489, 4565

\bibitem[{{Kundu} {et~al.}(2019{\natexlab{b}}){Kundu}, {Minniti}, \&
  {Singh}}]{kundu19}
{Kundu}, R., {Minniti}, D., \& {Singh}, H.~P. 2019{\natexlab{b}}, MNRAS, 483,
  1737

\bibitem[{{Kundu} {et~al.}(2021){Kundu}, {Navarrete}, {Fern{\'a}ndez-Trincado},
  {Minniti}, {Singh}, {Sbordone}, {Piatti}, \& {Reyl{\'e}}}]{Kundu20}
{Kundu}, R., {Navarrete}, C., {Fern{\'a}ndez-Trincado}, J.~G., {et~al.} 2021,
  A\&A, 645, A116

\bibitem[{{K{\"u}pper} {et~al.}(2010){K{\"u}pper}, {Kroupa}, {Baumgardt}, \&
  {Heggie}}]{Kupper10}
{K{\"u}pper}, A. H.~W., {Kroupa}, P., {Baumgardt}, H., \& {Heggie}, D.~C. 2010,
  MNRAS, 407, 2241

\bibitem[{{Kuzma} {et~al.}(2015){Kuzma}, {Da Costa}, {Keller}, \&
  {Maunder}}]{kuzma15}
{Kuzma}, P.~B., {Da Costa}, G.~S., {Keller}, S.~C., \& {Maunder}, E. 2015,
  MNRAS, 446, 3297

\bibitem[{Kuzma {et~al.}(2017)Kuzma, Da~Costa, \& Mackey}]{kuzma17}
Kuzma, P.~B., Da~Costa, G.~S., \& Mackey, A.~D. 2017, MNRAS, 473, 2881

\bibitem[{{Kuzma} {et~al.}(2016){Kuzma}, {Da Costa}, {Mackey}, \&
  {Roderick}}]{kuzma16}
{Kuzma}, P.~B., {Da Costa}, G.~S., {Mackey}, A.~D., \& {Roderick}, T.~A. 2016,
  MNRAS, 461, 3639

\bibitem[{{Law} \& {Majewski}(2010{\natexlab{a}})}]{law10}
{Law}, D.~R. \& {Majewski}, S.~R. 2010{\natexlab{a}}, ApJ, 718, 1128

\bibitem[{{Law} \& {Majewski}(2010{\natexlab{b}})}]{law10b}
{Law}, D.~R. \& {Majewski}, S.~R. 2010{\natexlab{b}}, ApJ, 714, 229

\bibitem[{{Law} \& {Majewski}(2016)}]{lawbook}
{Law}, D.~R. \& {Majewski}, S.~R. 2016, Astrophysics and Space Science Library,
  Vol. 420, {The Sagittarius Dwarf Tidal Stream(s)}, ed. H.~J. {Newberg} \&
  J.~L. {Carlin}, 31

\bibitem[{{Leon} {et~al.}(2000){Leon}, {Meylan}, \& {Combes}}]{leon00}
{Leon}, S., {Meylan}, G., \& {Combes}, F. 2000, A\&A, 359, 907

\bibitem[{{Lindegren} {et~al.}(2018){Lindegren}, {Hern{\'a}ndez}, {Bombrun},
  {Klioner}, {Bastian}, {Ramos-Lerate}, {de Torres}, {Steidelm{\"u}ller},
  {Stephenson}, {Hobbs}, {Lammers}, {Biermann}, {Geyer}, {Hilger}, {Michalik},
  {Stampa}, {McMillan}, {Casta{\~n}eda}, {Clotet}, {Comoretto}, {Davidson},
  {Fabricius}, {Gracia}, {Hambly}, {Hutton}, {Mora}, {Portell}, {van Leeuwen},
  {Abbas}, {Abreu}, {Altmann}, {Andrei}, {Anglada}, {Balaguer-N{\'u}{\~n}ez},
  {Barache}, {Becciani}, {Bertone}, {Bianchi}, {Bouquillon}, {Bourda},
  {Br{\"u}semeister}, {Bucciarelli}, {Busonero}, {Buzzi}, {Cancelliere},
  {Carlucci}, {Charlot}, {Cheek}, {Crosta}, {Crowley}, {de Bruijne}, {de
  Felice}, {Drimmel}, {Esquej}, {Fienga}, {Fraile}, {Gai}, {Garralda},
  {Gonz{\'a}lez-Vidal}, {Guerra}, {Hauser}, {Hofmann}, {Holl}, {Jordan},
  {Lattanzi}, {Lenhardt}, {Liao}, {Licata}, {Lister}, {L{\"o}ffler},
  {Marchant}, {Martin-Fleitas}, {Messineo}, {Mignard}, {Morbidelli}, {Poggio},
  {Riva}, {Rowell}, {Salguero}, {Sarasso}, {Sciacca}, {Siddiqui}, {Smart},
  {Spagna}, {Steele}, {Taris}, {Torra}, {van Elteren}, {van Reeven}, \&
  {Vecchiato}}]{Lindegren18}
{Lindegren}, L., {Hern{\'a}ndez}, J., {Bombrun}, A., {et~al.} 2018, A\&A, 616,
  A2

\bibitem[{Mackey \& Van Den~Bergh(2005)}]{mackey05}
Mackey, A.~D. \& Van Den~Bergh, S. 2005, MNRAS, 360, 631–645

\bibitem[{{Majewski} {et~al.}(2017){Majewski}, {Schiavon}, {Frinchaboy},
  {Allende Prieto}, {Barkhouser}, {Bizyaev}, {Blank}, {Brunner}, {Burton},
  {Carrera}, {Chojnowski}, {Cunha}, {Epstein}, {Fitzgerald}, {Garc{\'\i}a
  P{\'e}rez}, {Hearty}, {Henderson}, {Holtzman}, {Johnson}, {Lam}, {Lawler},
  {Maseman}, {M{\'e}sz{\'a}ros}, {Nelson}, {Nguyen}, {Nidever}, {Pinsonneault},
  {Shetrone}, {Smee}, {Smith}, {Stolberg}, {Skrutskie}, {Walker}, {Wilson},
  {Zasowski}, {Anders}, {Basu}, {Beland}, {Blanton}, {Bovy}, {Brownstein},
  {Carlberg}, {Chaplin}, {Chiappini}, {Eisenstein}, {Elsworth}, {Feuillet},
  {Fleming}, {Galbraith-Frew}, {Garc{\'\i}a}, {Garc{\'\i}a-Hern{\'a}ndez},
  {Gillespie}, {Girardi}, {Gunn}, {Hasselquist}, {Hayden}, {Hekker}, {Ivans},
  {Kinemuchi}, {Klaene}, {Mahadevan}, {Mathur}, {Mosser}, {Muna}, {Munn},
  {Nichol}, {O'Connell}, {Parejko}, {Robin}, {Rocha-Pinto}, {Schultheis},
  {Serenelli}, {Shane}, {Silva Aguirre}, {Sobeck}, {Thompson}, {Troup},
  {Weinberg}, \& {Zamora}}]{apogee}
{Majewski}, S.~R., {Schiavon}, R.~P., {Frinchaboy}, P.~M., {et~al.} 2017, AJ,
  154, 94

\bibitem[{{Majewski} {et~al.}(2003){Majewski}, {Skrutskie}, {Weinberg}, \&
  {Ostheimer}}]{Majewski03}
{Majewski}, S.~R., {Skrutskie}, M.~F., {Weinberg}, M.~D., \& {Ostheimer}, J.~C.
  2003, ApJ, 599, 1082

\bibitem[{{Malhan} {et~al.}(2018){Malhan}, {Ibata}, \& {Martin}}]{Malhan18b}
{Malhan}, K., {Ibata}, R.~A., \& {Martin}, N.~F. 2018, MNRAS, 481, 3442

\bibitem[{{Malhan} {et~al.}(2021){Malhan}, {Yuan}, {Ibata}, {Arentsen},
  {Bellazzini}, \& {Martin}}]{Malhan21}
{Malhan}, K., {Yuan}, Z., {Ibata}, R., {et~al.} 2021, arXiv e-prints,
  arXiv:2104.09523

\bibitem[{Marigo {et~al.}(2017)Marigo, Girardi, Bressan, Rosenfield, Aringer,
  Chen, Dussin, Nanni, Pastorelli, Rodrigues, Trabucchi, Bladh, Dalcanton,
  Groenewegen, Montalb{\'{a}}n, \& Wood}]{Marigo17}
Marigo, P., Girardi, L., Bressan, A., {et~al.} 2017, The Astrophysical Journal,
  835, 77

\bibitem[{{Mart{\'\i}nez-Delgado} {et~al.}(2001){Mart{\'\i}nez-Delgado},
  {Aparicio}, {G{\'o}mez-Flechoso}, \& {Carrera}}]{Delgado02}
{Mart{\'\i}nez-Delgado}, D., {Aparicio}, A., {G{\'o}mez-Flechoso}, M.~{\'A}.,
  \& {Carrera}, R. 2001, ApJl, 549, L199

\bibitem[{{Mart{\'\i}nez-Delgado} {et~al.}(2002){Mart{\'\i}nez-Delgado},
  {Zinn}, {Carrera}, \& {Gallart}}]{martnez02}
{Mart{\'\i}nez-Delgado}, D., {Zinn}, R., {Carrera}, R., \& {Gallart}, C. 2002,
  ApJl, 573, L19

\bibitem[{{Massari} {et~al.}(2019){Massari}, {Koppelman}, \&
  {Helmi}}]{Massari19}
{Massari}, D., {Koppelman}, H.~H., \& {Helmi}, A. 2019, A\&A, 630, L4

\bibitem[{{Minniti} {et~al.}(2021{\natexlab{a}}){Minniti}, {G{\'o}mez},
  {Alonso-Garc{\'\i}a}, {Saito}, \& {Garro}}]{dante21a}
{Minniti}, D., {G{\'o}mez}, M., {Alonso-Garc{\'\i}a}, J., {Saito}, R.~K., \&
  {Garro}, E.~R. 2021{\natexlab{a}}, A\&A, 650, L12

\bibitem[{{Minniti} {et~al.}(2021{\natexlab{b}}){Minniti}, {Ripepi},
  {Fern{\'a}ndez-Trincado}, {Alonso-Garc{\'\i}a}, {Smith}, {Lucas},
  {G{\'o}mez}, {Pullen}, {Garro}, {Vivanco C{\'a}diz}, {Hempel}, {Rejkuba},
  {Saito}, {Palma}, {Clari{\'a}}, {Gregg}, \& {Majaess}}]{dante21}
{Minniti}, D., {Ripepi}, V., {Fern{\'a}ndez-Trincado}, J.~G., {et~al.}
  2021{\natexlab{b}}, A\&A, 647, L4

\bibitem[{Monaco {et~al.}(2005)Monaco, Bellazzini, Ferraro, \&
  Pancino}]{monaco05}
Monaco, L., Bellazzini, M., Ferraro, F.~R., \& Pancino, E. 2005, MNRAS, 356,
  1396

\bibitem[{{Mottini} {et~al.}(2008){Mottini}, {Wallerstein}, \&
  {McWilliam}}]{Mottini08}
{Mottini}, M., {Wallerstein}, G., \& {McWilliam}, A. 2008, AJ, 136, 614

\bibitem[{{Musella} {et~al.}(2018){Musella}, {Di Criscienzo}, {Marconi},
  {Raimondo}, {Ripepi}, {Cignoni}, {Bono}, {Brocato}, {Dall'Ora}, {Ferraro},
  {Grado}, {Iannicola}, {Limatola}, {Molinaro}, {Moretti}, {Stetson},
  {Capaccioli}, {Cioni}, {Getman}, \& {Schipani}}]{Musella18}
{Musella}, I., {Di Criscienzo}, M., {Marconi}, M., {et~al.} 2018, MNRAS, 473,
  3062

\bibitem[{{Navarrete} {et~al.}(2017){Navarrete}, {Belokurov}, {Koposov},
  {Irwin}, {Catelan}, {Duffau}, \& {Drake}}]{Navarrete17}
{Navarrete}, C., {Belokurov}, V., {Koposov}, S.~E., {et~al.} 2017, MNRAS, 467,
  1329

\bibitem[{{Newberg} {et~al.}(2002){Newberg}, {Yanny}, {Rockosi}, {Grebel},
  {Rix}, {Brinkmann}, {Csabai}, {Hennessy}, {Hindsley}, {Ibata}, {Ivezi{\'c}},
  {Lamb}, {Nash}, {Odenkirchen}, {Rave}, {Schneider}, {Smith}, {Stolte}, \&
  {York}}]{Newberg02}
{Newberg}, H.~J., {Yanny}, B., {Rockosi}, C., {et~al.} 2002, ApJ, 569, 245

\bibitem[{{Palau} \& {Miralda-Escud{\'e}}(2019)}]{Palau19}
{Palau}, C.~G. \& {Miralda-Escud{\'e}}, J. 2019, MNRAS, 488, 1535

\bibitem[{{Pancino} {et~al.}(2017){Pancino}, {Bellazzini}, {Giuffrida}, \&
  {Marinoni}}]{Pancino17}
{Pancino}, E., {Bellazzini}, M., {Giuffrida}, G., \& {Marinoni}, S. 2017,
  MNRAS, 467, 412

\bibitem[{Pedregosa {et~al.}(2011)Pedregosa, Varoquaux, Gramfort, Michel,
  Thirion, Grisel, Blondel, Prettenhofer, Weiss, Dubourg, Vanderplas, Passos,
  Cournapeau, Brucher, Perrot, \& Duchesnay}]{sklearn}
Pedregosa, F., Varoquaux, G., Gramfort, A., {et~al.} 2011, Journal of Machine
  Learning Research, 12, 2825

\bibitem[{{Piatti}(2021)}]{piatti21}
{Piatti}, A.~E. 2021, \aj, 162, 261

\bibitem[{{Price-Whelan}(2017)}]{galacoor}
{Price-Whelan}, A.~M. 2017, The Journal of Open Source Software, 2, 388

\bibitem[{{Puzia} {et~al.}(2005){Puzia}, {Kissler-Patig}, {Thomas}, {Maraston},
  {Saglia}, {Bender}, {Goudfrooij}, \& {Hempel}}]{Puzia05}
{Puzia}, T.~H., {Kissler-Patig}, M., {Thomas}, D., {et~al.} 2005, A\&A, 439,
  997

\bibitem[{Riello {et~al.}(2021)Riello, De~Angeli, Evans, Montegriffo, Carrasco,
  Busso, Palaversa, Burgess, Diener, Davidson, \& et~al.}]{Riello21}
Riello, M., De~Angeli, F., Evans, D.~W., {et~al.} 2021, A\&A, 649, A3

\bibitem[{{Ripepi} {et~al.}(2007){Ripepi}, {Clementini}, {Di Criscienzo},
  {Greco}, {Dall'Ora}, {Federici}, {Di Fabrizio}, {Musella}, {Marconi},
  {Baldacci}, \& {Maio}}]{Ripepi07}
{Ripepi}, V., {Clementini}, G., {Di Criscienzo}, M., {et~al.} 2007, ApJl, 667,
  L61

\bibitem[{{Rodriguez-Gomez} {et~al.}(2016){Rodriguez-Gomez}, {Pillepich},
  {Sales}, {Genel}, {Vogelsberger}, {Zhu}, {Wellons}, {Nelson}, {Torrey},
  {Springel}, {Ma}, \& {Hernquist}}]{Gomez16}
{Rodriguez-Gomez}, V., {Pillepich}, A., {Sales}, L.~V., {et~al.} 2016, MNRAS,
  458, 2371

\bibitem[{{Salinas} {et~al.}(2012){Salinas}, {J{\'\i}lkov{\'a}}, {Carraro},
  {Catelan}, \& {Amigo}}]{Salinas12}
{Salinas}, R., {J{\'\i}lkov{\'a}}, L., {Carraro}, G., {Catelan}, M., \&
  {Amigo}, P. 2012, MNRAS, 421, 960

\bibitem[{{Sariya} \& {Yadav}(2015)}]{Sariya15}
{Sariya}, D.~P. \& {Yadav}, R.~K.~S. 2015, A\&A, 584, A59

\bibitem[{{Sbordone} {et~al.}(2007{\natexlab{a}}){Sbordone}, {Bonifacio},
  {Buonanno}, {Marconi}, {Monaco}, \& {Zaggia}}]{Sbordone07}
{Sbordone}, L., {Bonifacio}, P., {Buonanno}, R., {et~al.} 2007{\natexlab{a}},
  A\&A, 465, 815

\bibitem[{{Sbordone} {et~al.}(2007{\natexlab{b}}){Sbordone}, {Bonifacio},
  {Giuffrida}, {Marconi}, {Monaco}, \& {Zaggia}}]{luca07}
{Sbordone}, L., {Bonifacio}, P., {Giuffrida}, G., {et~al.} 2007{\natexlab{b}},
  in Galaxy Evolution across the Hubble Time, ed. F.~{Combes} \&
  J.~{Palou{\v{s}}}, Vol. 235, 330--330

\bibitem[{{Sbordone} {et~al.}(2005){Sbordone}, {Bonifacio}, {Marconi},
  {Buonanno}, \& {Zaggia}}]{luca05}
{Sbordone}, L., {Bonifacio}, P., {Marconi}, G., {Buonanno}, R., \& {Zaggia}, S.
  2005, A\&A, 437, 905

\bibitem[{{Sbordone} {et~al.}(2003){Sbordone}, {Marconi}, {Bonifacio}, \&
  {Buonanno}}]{luca03}
{Sbordone}, L., {Marconi}, G., {Bonifacio}, P., \& {Buonanno}, R. 2003, in IAU
  Joint Discussion, Vol.~25, IAU Joint Discussion, E36

\bibitem[{{Sbordone} {et~al.}(2015){Sbordone}, {Monaco}, {Moni Bidin},
  {Bonifacio}, {Villanova}, {Bellazzini}, {Ibata}, {Chiba}, {Geisler},
  {Caffau}, \& {Duffau}}]{luca15}
{Sbordone}, L., {Monaco}, L., {Moni Bidin}, C., {et~al.} 2015, A\&A, 579, A104

\bibitem[{{Schlafly} \& {Finkbeiner}(2011)}]{Schlafly11}
{Schlafly}, E.~F. \& {Finkbeiner}, D.~P. 2011, ApJ, 737, 103

\bibitem[{{Schwarz}(1978)}]{Schwarz78}
{Schwarz}, G. 1978, Annals of Statistics, 6, 461

\bibitem[{Sharina {et~al.}(2013)Sharina, Shimansky, \& Davoust}]{Sharina13}
Sharina, M.~E., Shimansky, V.~V., \& Davoust, E. 2013, Astronomy Reports, 57,
  410–422

\bibitem[{{Siegel} {et~al.}(2007){Siegel}, {Dotter}, {Majewski}, {Sarajedini},
  {Chaboyer}, {Nidever}, {Anderson}, {Mar{\'\i}n-Franch}, {Rosenberg}, {Bedin},
  {Aparicio}, {King}, {Piotto}, \& {Reid}}]{Siegel07}
{Siegel}, M.~H., {Dotter}, A., {Majewski}, S.~R., {et~al.} 2007, ApJl, 667, L57

\bibitem[{{Sohn} {et~al.}(2018){Sohn}, {Watkins}, {Fardal}, {van der Marel},
  {Deason}, {Besla}, \& {Bellini}}]{Sohn18}
{Sohn}, S.~T., {Watkins}, L.~L., {Fardal}, M.~A., {et~al.} 2018, ApJ, 862, 52

\bibitem[{{Tang} {et~al.}(2018){Tang}, {Fern{\'a}ndez-Trincado}, {Geisler},
  {Zamora}, {M{\'e}sz{\'a}ros}, {Masseron}, {Cohen},
  {Garc{\'\i}a-Hern{\'a}ndez}, {Dell'Agli}, {Beers}, {Schiavon}, {Sohn},
  {Hasselquist}, {Robin}, {Shetrone}, {Majewski}, {Villanova}, {Schiappacasse
  Ulloa}, {Lane}, {Minnti}, {Roman-Lopes}, {Almeida}, \& {Moreno}}]{tang18}
{Tang}, B., {Fern{\'a}ndez-Trincado}, J.~G., {Geisler}, D., {et~al.} 2018, ApJ,
  855, 38

\bibitem[{Valcarce {et~al.}(2012)Valcarce, Catelan, \& Sweigart}]{Valcarce12}
Valcarce, A. A.~R., Catelan, M., \& Sweigart, A.~V. 2012, A\&A, 547, A5

\bibitem[{{Valcarce} {et~al.}(2012){Valcarce}, {Catelan}, \& {Sweigart}}]{zahb}
{Valcarce}, A.~A.~R., {Catelan}, M., \& {Sweigart}, A.~V. 2012, A\&A, 547, A5

\bibitem[{{Vasiliev} \& {Baumgardt}(2021)}]{GC}
{Vasiliev}, E. \& {Baumgardt}, H. 2021, \mnras, 505, 5978

\bibitem[{{Villanova} {et~al.}(2016){Villanova}, {Monaco}, {Moni Bidin}, \&
  {Assmann}}]{Villanova16}
{Villanova}, S., {Monaco}, L., {Moni Bidin}, C., \& {Assmann}, P. 2016, MNRAS,
  460, 2351–2359

\bibitem[{Virtanen {et~al.}(2020)Virtanen, Gommers, Oliphant, Haberland, Reddy,
  Cournapeau, Burovski, Peterson, Weckesser, Bright, {van der Walt}, Brett,
  Wilson, Millman, Mayorov, Nelson, Jones, Kern, Larson, Carey, Polat, Feng,
  Moore, {VanderPlas}, Laxalde, Perktold, Cimrman, Henriksen, Quintero, Harris,
  Archibald, Ribeiro, Pedregosa, {van Mulbregt}, \& {SciPy 1.0
  Contributors}}]{scipy}
Virtanen, P., Gommers, R., Oliphant, T.~E., {et~al.} 2020, Nature Methods, 17,
  261

\bibitem[{{Yuan} {et~al.}(2020){Yuan}, {Chang}, {Beers}, \& {Huang}}]{Yuan20}
{Yuan}, Z., {Chang}, J., {Beers}, T.~C., \& {Huang}, Y. 2020, ApJl, 898, L37

\end{thebibliography}


\begin{thebibliography}{29}
\expandafter\ifx\csname natexlab\endcsname\relax\def\natexlab#1{#1}\fi

\bibitem[{{Carretta} {et~al.}(2009){Carretta}, {Bragaglia}, {Gratton}, \&
  {Lucatello}}]{Carretta09}
{Carretta}, E., {Bragaglia}, A., {Gratton}, R., \& {Lucatello}, S. 2009, AAP,
  505, 139

\bibitem[{{Cool} \& {Bolton}(2002)}]{cool02}
{Cool}, A.~M. \& {Bolton}, A.~S. 2002, in Astronomical Society of the Pacific
  Conference Series, Vol. 263, Stellar Collisions, Mergers and their
  Consequences, ed. M.~M. {Shara}, 163

\bibitem[{{Cudworth} \& {Hanson}(1993)}]{Cudworth93}
{Cudworth}, K.~M. \& {Hanson}, R.~B. 1993, AJ, 105, 168

\bibitem[{{de Marchi} \& {Paresce}(1994)}]{Marchi94}
{de Marchi}, G. \& {Paresce}, F. 1994, AAP, 281, L13

\bibitem[{{Dinescu} {et~al.}(1999){Dinescu}, {Girard}, \& {van
  Altena}}]{dana99}
{Dinescu}, D.~I., {Girard}, T.~M., \& {van Altena}, W.~F. 1999, AJ, 117, 1792

\bibitem[{{Evans} {et~al.}(2018){Evans}, {Riello}, {De Angeli}, {Carrasco},
  {Montegriffo}, {Fabricius}, {Jordi}, {Palaversa}, {Diener}, {Busso},
  {Cacciari}, {van Leeuwen}, {Burgess}, {Davidson}, {Harrison}, {Hodgkin},
  {Pancino}, {Richards}, {Altavilla}, {Balaguer-N{\'u}{\~n}ez}, {Barstow},
  {Bellazzini}, {Brown}, {Castellani}, {Cocozza}, {De Luise}, {Delgado},
  {Ducourant}, {Galleti}, {Gilmore}, {Giuffrida}, {Holl}, {Kewley}, {Koposov},
  {Marinoni}, {Marrese}, {Osborne}, {Piersimoni}, {Portell}, {Pulone},
  {Ragaini}, {Sanna}, {Terrett}, {Walton}, {Wevers}, \&
  {Wyrzykowski}}]{evans18}
{Evans}, D.~W., {Riello}, M., {De Angeli}, F., {et~al.} 2018, AAP, 616, A4

\bibitem[{{Fern{\'a}ndez-Trincado} {et~al.}(2018){Fern{\'a}ndez-Trincado},
  {Vega Neme}, {Vieira}, {G{\'o}mez-L{\'o}pez}, \& {Verdugo}}]{jose18tidal}
{Fern{\'a}ndez-Trincado}, J.~G., {Vega Neme}, L.~R., {Vieira}, K.,
  {G{\'o}mez-L{\'o}pez}, J.~A., \& {Verdugo}, T. 2018, in Terceras Jornadas de
  Astrof\&iacute;sica Estelar, 126--129

\bibitem[{Ferraro {et~al.}(1997)Ferraro, Carretta, Bragaglia, Renzini, \&
  Ortolani}]{Ferraro97}
Ferraro, F.~R., Carretta, E., Bragaglia, A., Renzini, A., \& Ortolani, S. 1997,
  MNRAS, 286, 1012

\bibitem[{{Gaia Collaboration} {et~al.}(2018){Gaia Collaboration}, {Brown},
  {Vallenari}, {Prusti}, {de Bruijne}, {Babusiaux}, \&
  {Bailer-Jones}}]{gaiadr2}
{Gaia Collaboration}, {Brown}, A.~G.~A., {Vallenari}, A., {et~al.} 2018, ArXiv
  e-prints [\eprint[arXiv]{1804.09365}]

\bibitem[{{Gnedin} {et~al.}(1999){Gnedin}, {Lee}, \& {Ostriker}}]{oleg99}
{Gnedin}, O.~Y., {Lee}, H.~M., \& {Ostriker}, J.~P. 1999, ApJ, 522, 935

\bibitem[{{Goldsbury} {et~al.}(2013){Goldsbury}, {Heyl}, \&
  {Richer}}]{Goldsbury13}
{Goldsbury}, R., {Heyl}, J., \& {Richer}, H. 2013, ApJ, 778, 57

\bibitem[{{Gratton} {et~al.}(2003){Gratton}, {Bragaglia}, {Carretta},
  {Clementini}, {Desidera}, {Grundahl}, \& {Lucatello}}]{Gratton03}
{Gratton}, R.~G., {Bragaglia}, A., {Carretta}, E., {et~al.} 2003, A\&A, 408,
  529

\bibitem[{{Harris}(1996)}]{harris96}
{Harris}, W.~E. 1996, \aj, 112, 1487

\bibitem[{{Heyl} {et~al.}(2012){Heyl}, {Richer}, {Anderson}, {Fahlman},
  {Dotter}, {Hurley}, {Kalirai}, {Rich}, {Shara}, {Stetson}, {Woodley}, \&
  {Zurek}}]{heyl12}
{Heyl}, J.~S., {Richer}, H., {Anderson}, J., {et~al.} 2012, ApJ, 761, 51

\bibitem[{{Husser} {et~al.}(2016){Husser}, {Kamann}, {Dreizler}, {Wendt},
  {Wulff}, {Bacon}, {Wisotzki}, {Brinchmann}, {Weilbacher}, {Roth}, \&
  {Monreal-Ibero}}]{Husser16}
{Husser}, T.-O., {Kamann}, S., {Dreizler}, S., {et~al.} 2016, A\&A, 588, A148

\bibitem[{King {et~al.}(1995)King, Sosin, \& Cool}]{King95}
King, I.~R., Sosin, C., \& Cool, A.~M. 1995, ApJ, 452

\bibitem[{{Kundu} {et~al.}(2019{\natexlab{a}}){Kundu},
  {Fern{\'a}ndez-Trincado}, {Minniti}, {Singh}, {Moreno}, {Reyl{\'e}}, {Robin},
  \& {Soto}}]{kundu19}
{Kundu}, R., {Fern{\'a}ndez-Trincado}, J.~G., {Minniti}, D., {et~al.}
  2019{\natexlab{a}}, \mnras, 489, 4565

\bibitem[{{Kundu} {et~al.}(2019{\natexlab{b}}){Kundu}, {Minniti}, \&
  {Singh}}]{kundu19a}
{Kundu}, R., {Minniti}, D., \& {Singh}, H.~P. 2019{\natexlab{b}}, MNRAS, 483,
  1737

\bibitem[{{Leon} {et~al.}(2000){Leon}, {Meylan}, \& {Combes}}]{Leon00}
{Leon}, S., {Meylan}, G., \& {Combes}, F. 2000, A\&A, 359, 907

\bibitem[{{Martinazzi} {et~al.}(2014){Martinazzi}, {Pieres}, {Kepler}, {Costa},
  {Bonatto}, \& {Bica}}]{Martinazzi14}
{Martinazzi}, E., {Pieres}, A., {Kepler}, S.~O., {et~al.} 2014, MNRAS, 442,
  3105

\bibitem[{{Meszaros et. al.}(2019)}]{Meszaros19}
{Meszaros et. al.} 2019, in preperation

\bibitem[{{Milone} {et~al.}(2012){Milone}, {Marino}, {Piotto}, {Bedin},
  {Anderson}, {Aparicio}, {Cassisi}, \& {Rich}}]{Milone12}
{Milone}, A.~P., {Marino}, A.~F., {Piotto}, G., {et~al.} 2012, ApJ, 745, 27

\bibitem[{{Minniti} {et~al.}(2018){Minniti}, {Fern{\'a}ndez-Trincado},
  {Ripepi}, {Alonso-Garc{\'\i}a}, {Contreras Ramos}, \& {Marconi}}]{dante18}
{Minniti}, D., {Fern{\'a}ndez-Trincado}, J.~G., {Ripepi}, V., {et~al.} 2018,
  ApJl, 869, L10

\bibitem[{{Minniti} {et~al.}(1993){Minniti}, {Geisler}, {Peterson}, \&
  {Claria}}]{miniti93}
{Minniti}, D., {Geisler}, D., {Peterson}, R.~C., \& {Claria}, J.~J. 1993, ApJ,
  413, 548

\bibitem[{{Parker} {et~al.}(2016){Parker}, {Goodwin}, {Wright}, {Meyer}, \&
  {Quanz}}]{Parker16}
{Parker}, R.~J., {Goodwin}, S.~P., {Wright}, N.~J., {Meyer}, M.~R., \& {Quanz},
  S.~P. 2016, MNRAS, 459, L119

\bibitem[{{Pecaut} \& {Mamajek}(2013)}]{pecaut13}
{Pecaut}, M.~J. \& {Mamajek}, E.~E. 2013, ApJs, 208, 9

\bibitem[{{Pittordis} \& {Sutherland}(2019)}]{pittordis19}
{Pittordis}, C. \& {Sutherland}, W. 2019, MNRAS, 488, 4740

\bibitem[{{Sosin} {et~al.}(1994){Sosin}, {King}, \& {Cool}}]{Sosin94}
{Sosin}, C., {King}, I.~R., \& {Cool}, A.~M. 1994, in American Astronomical
  Society Meeting Abstracts, Vol. 185, 103.09

\bibitem[{{Vasiliev}(2019)}]{GC}
{Vasiliev}, E. 2019, MNRAS, 484, 2832

\end{thebibliography}


\begin{appendix}

\section{Proper motion estimates} \label{sec:appendix}

In this Appendix, we compare the results obtained in Section~\ref{sec:ppm} with those reported in 
\citet[hereinafter VB21]{GC}. Table~\ref{tab:ppm} contains the mean PMs, observed dispersions, and intrinsic dispersions. To estimate the errors, the Gaussian Mixture Model fit was run 5000 times, randomly sampling 90\% of the original sample of stars for each cluster, adopting as errors for each value the standard deviation of its distribution.
\begin{table*}[htbp]
        \centering
        \caption{Mean PMs, dispersions and intrinsic PM dispersions for the clusters studied in this work.}
        \label{tab:ppm}
        \begin{tabular}{cccccc} 
\hline
Cluster & $\mu_{\alpha,0}$ &  $\mu_{\delta}$    & $\sigma_{\mu_{\alpha,0}}$ & $\sigma_{\mu_{\delta}}$ & $\sigma_{\mu}$ \\
        & (mas yr$^{-1}$)  &  (mas yr$^{-1}$)   & (mas yr$^{-1}$)           &    (mas yr$^{-1}$)     & (mas yr$^{-1}$) \\
\hline
Arp 2    & --2.37 $\pm$ 0.04 & --1.51 $\pm$ 0.06 & 0.30 $\pm$ 0.03           & 0.18 $\pm$ 0.03       & 0.201 \\
Terzan 8 & --2.48 $\pm$ 0.03 & --1.58 $\pm$ 0.02 & 0.17 $\pm$ 0.08           & 0.12 $\pm$ 0.07       & 0.051 \\
M 54     & --2.68 $\pm$ 0.01 & --1.39 $\pm$ 0.01 & 0.20 $\pm$ 0.01           & 0.16 $\pm$ 0.01       & 0.189 \\ 
Terzan 7 & --2.97 $\pm$ 0.02 & --1.65 $\pm$ 0.02 & 0.16 $\pm$ 0.05           & 0.17 $\pm$ 0.08       & 0.098 \\
NGC 5634 & --1.70 $\pm$ 0.01 & --1.47 $\pm$ 0.01 & 0.16 $\pm$ 0.03           & 0.13 $\pm$ 0.02       & 0.065 \\ 
NGC 2419 & --0.04 $\pm$ 0.02 & --0.52 $\pm$ 0.01 & 0.33 $\pm$ 0.02           & 0.21 $\pm$ 0.02       & 0.276 \\
NGC 4147 & --1.71 $\pm$ 0.02 & --2.08 $\pm$ 0.02 & 0.15 $\pm$ 0.02           & 0.13 $\pm$ 0.02       & 0.048 \\
Pal 12   & --3.22 $\pm$ 0.03 & --3.36 $\pm$ 0.03 & 0.24 $\pm$ 0.07           & 0.12 $\pm$ 0.07       & 0.199 \\
NGC 6284 & --3.19 $\pm$ 0.03 & --2.02 $\pm$ 0.09 & 0.24 $\pm$ 0.02           & 0.11 $\pm$ 0.03       & 0.095 \\
\hline
\end{tabular}
\end{table*}

\begin{figure}[h!]
    \centering
    \includegraphics[scale=0.45]{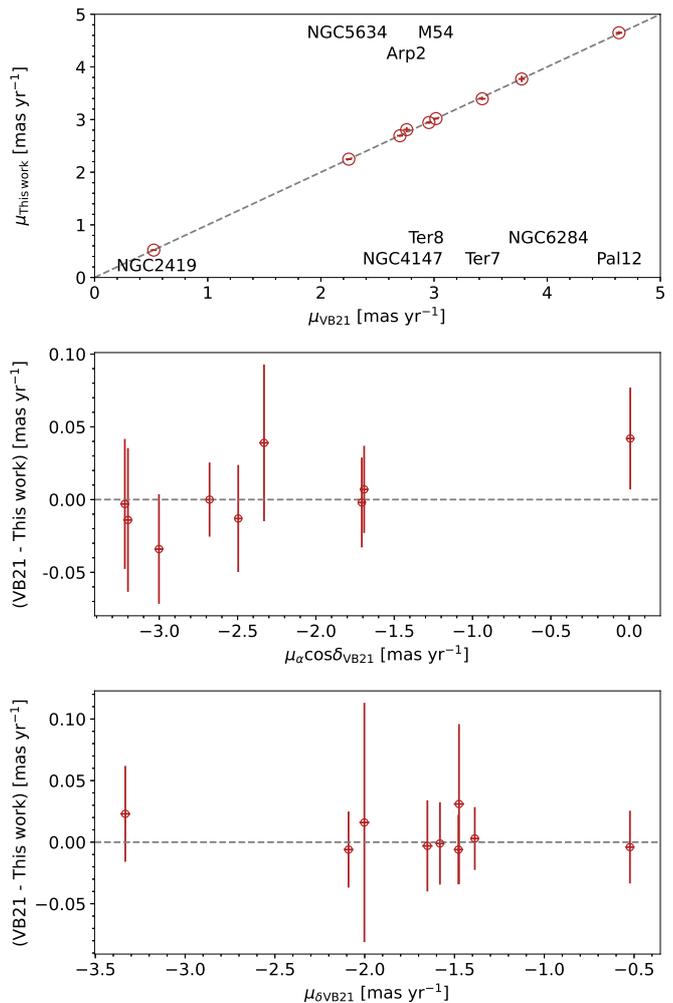}
    \caption{Comparison between the PMs derived in this work and those reported 
    in VB21. Top panel: Total PM from VB21 and this work. The 
    dashed line represents the 1:1 line. The middle and bottom panels show the difference between 
    VB21 and this work for $\mu_{\alpha,0}$ and $\mu_{\delta}$, respectively.}
    \label{fig:ppm_comp}
\end{figure}

Figure~\ref{fig:ppm_comp} shows the comparison between the mean PMs derived in this work and 
the values reported in VB21. There is a very good agreement between the values used in this work and 
the ones reported in the literature. The clusters with the largest discrepancies are Terzan 7, Arp 2, and NGC 2419. This can be explained due to the relatively low number of cluster stars ($\sim$100) in the case of Terzan 7 and the elongated PM distributions of Arp 2 and NGC 2419. 

In fact, Table~\ref{tab:ppm_field} shows the mean PM and its dispersions for the field populations fitted along the PM distribution of the cluster members. For most of the clusters, the dispersion of this broad component is more than 1 mas yr$^{-1}$ in the $\mu_{\delta}$ axis. It is worth mentioning that these values can change considerably depending on the stars used to do the fitting. In this work, only stars inside one $r_t$, brighter than G $=$ 19 mag, and having absolute PM values lower than 10 mas yr$^{-1}$ in each direction were used. Being inside one $r_t$ of each cluster centre means that the field population is less significant in number and, due to this, the fitted parameters are not as well constrained as the parameters for the cluster population. This is the case for NGC 5634, in which the field population has almost the same mean PMs as the clusters, but a larger dispersion. This is due to the very low number of field stars inside one $r_t$ (see Figure~\ref{PM}).

\begin{table*}[htbp]
        \centering
        \caption{Mean PM and dispersions for the field populations around each cluster, up to one $r_t$.}
        \label{tab:ppm_field}
        \begin{tabular}{ccccc} 
\hline
Field  & $\mu_{\alpha,0}$ &  $\mu_{\delta}$    & $\sigma_{\mu_{\alpha,0}}$ & $\sigma_{\mu_{\delta}}$ \\
population & (mas yr$^{-1}$)  &  (mas yr$^{-1}$)   & (mas yr$^{-1}$)           & (mas yr$^{-1}$)     \\
\hline
Arp 2    & --0.72 $\pm$ 0.28 & --3.81 $\pm$ 0.26 & 3.62 $\pm$ 0.56           & 3.10 $\pm$ 0.46      \\
Terzan 8 & --1.17 $\pm$ 0.22 & --3.16 $\pm$ 0.22 & 3.46 $\pm$ 1.08           & 3.18 $\pm$ 1.03      \\
M 54     & --1.05 $\pm$ 0.08 & --3.73 $\pm$ 0.08 & 3.01 $\pm$ 0.06           & 3.02 $\pm$ 0.06      \\ 
Terzan 7 & --0.78 $\pm$ 1.38 & --1.65 $\pm$ 1.31 & 2.06 $\pm$ 0.95           & 1.54 $\pm$ 0.84      \\
NGC 5634 & --1.76 $\pm$ 0.92 & --1.53 $\pm$ 0.83 & 1.07 $\pm$ 0.83           & 0.94 $\pm$ 0.73      \\ 
NGC 2419 & --0.60 $\pm$ 0.24 & --3.25 $\pm$ 0.26 & 2.77 $\pm$ 0.21           & 2.65 $\pm$ 0.15      \\
NGC 4147 & --6.35 $\pm$ 3.51 & --8.14 $\pm$ 3.09 & 9.03 $\pm$ 3.54           & 5.34 $\pm$ 1.99      \\
Pal 12   &   0.78 $\pm$ 2.17 & --4.23 $\pm$ 0.40 & 4.19 $\pm$ 1.03           & 3.03 $\pm$ 0.56      \\
NGC 6284 & --1.58 $\pm$ 0.53 & --2.31 $\pm$ 0.14 & 2.07 $\pm$ 0.34           & 1.91 $\pm$ 0.19      \\
\hline
\end{tabular}
\end{table*}

\begin{figure}[h!]
    \centering
    \includegraphics[scale=0.45]{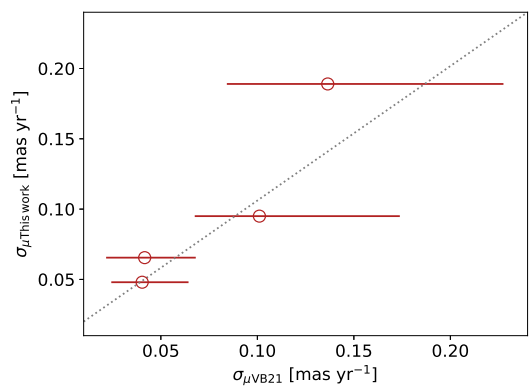}
    \caption{Comparison between the internal PM dispersion $\sigma_{\mu}$ at the centre of the cluster, 
    as reported in VB21, and the intrinsic PM dispersion obtained in this work, as explained in the text.}
    \label{fig:int_ppm}
\end{figure}

The dispersions reported in columns 4 and 5 in Table~\ref{tab:ppm} are the observed dispersions in the PM distribution of the clusters. The intrinsic dispersion of the PMs of the clusters (column six) was obtained 
subtracting, in quadrature, the mean observational errors in each component. Figure~\ref{fig:int_ppm} shows the PM dispersion at the 50th percentile, using as error bars the values at 5\% and 95\%, based on the PM dispersion profiles published by VB21 (for the four clusters included in this work that are also part of the former study, namely M 54, NGC 5634, NGC 4147, and NGC 6284) compared to the values found in this work. There is a relatively good agreement, although the values derived in this work tend to be slightly larger than the ones reported by VB21. Given the extensive modelling performed in the aforementioned study, the most likely explanation is that the dispersion values found in the fitting done in this work are overestimated. Therefore, the values for the intrinsic dispersion presented in Table~\ref{tab:par} should be considered as a rough estimate for those clusters without PM dispersion profiles in the work of VB21.

All the results presented above will need to be updated once the Gaia data release 3 data is available.

\section{Comparison fields for the inner-region clusters} \label{sec:appendix_fields}

\begin{figure*}
\begin{center}
\includegraphics[width=\textwidth, height=\textheight, keepaspectratio]{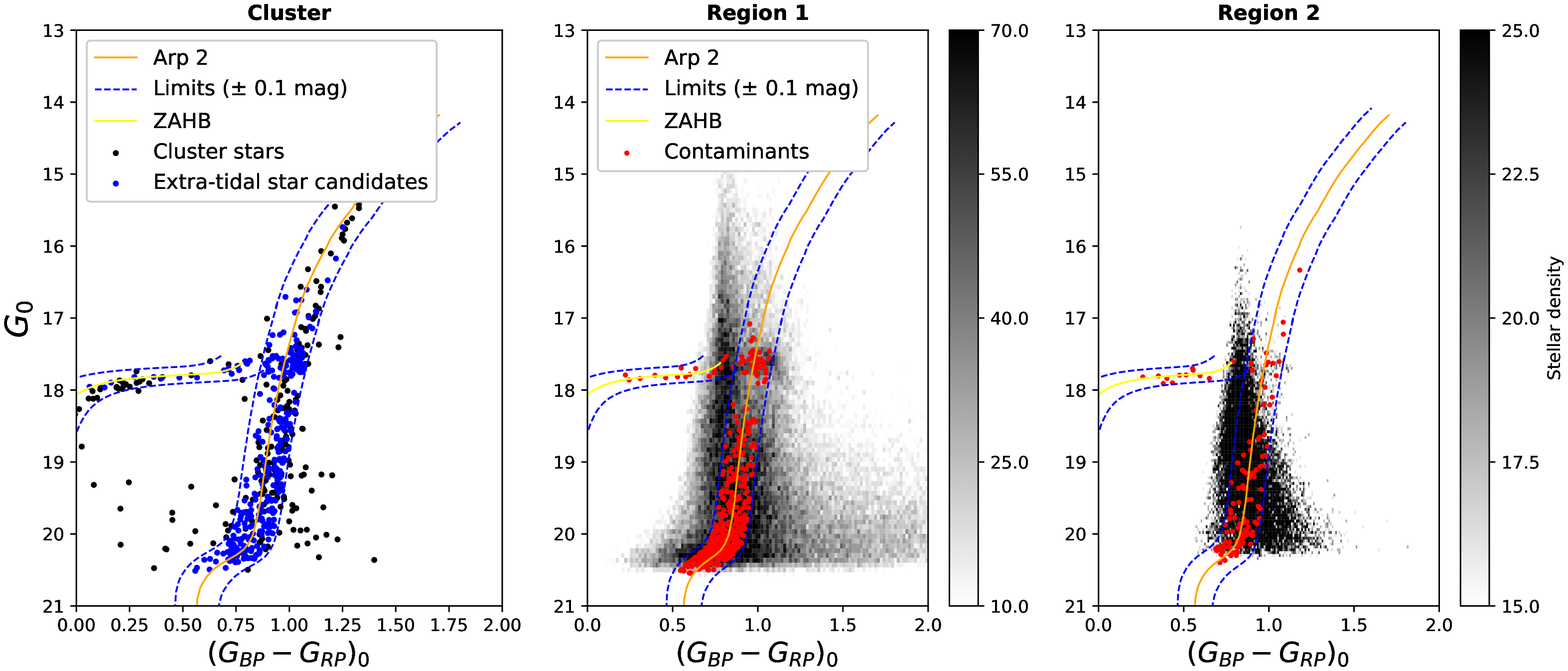}

\includegraphics[width=\textwidth, height=\textheight, keepaspectratio]{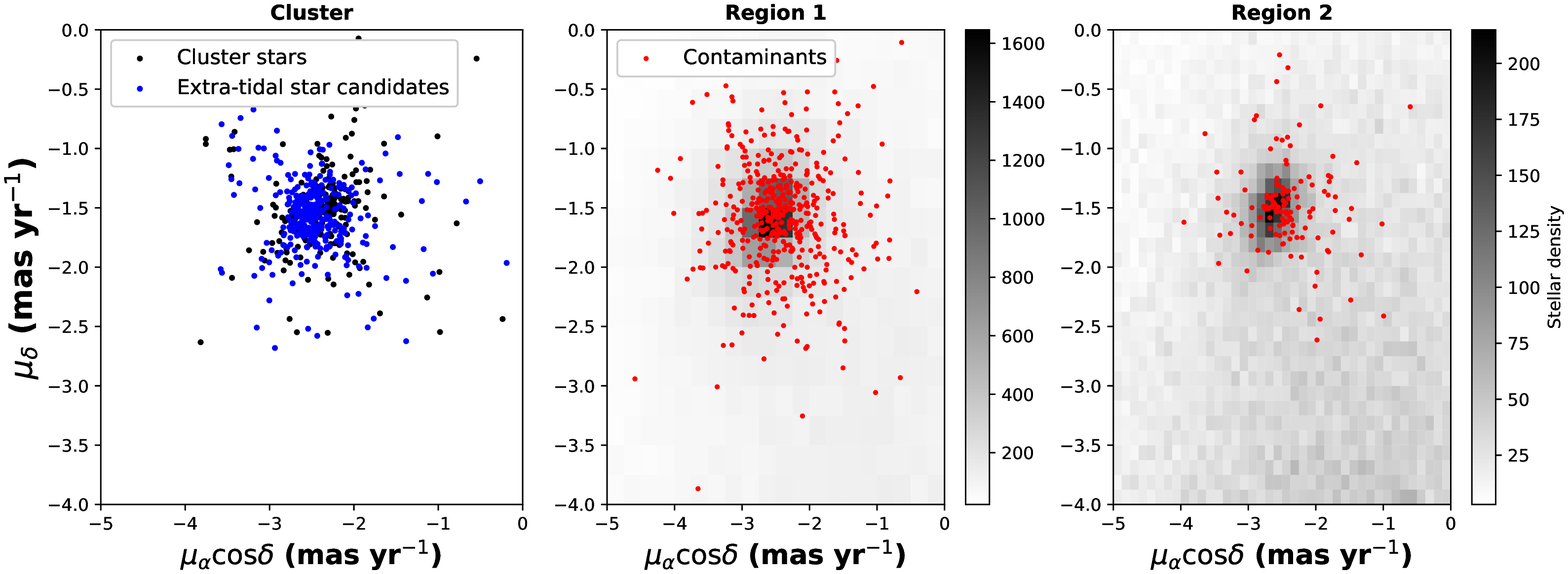}
\caption{Upper left panel shows the CMD of the cluster (black dots) overplotted with the extra-tidal stars (blue dots). Arp 2 isochrone and ZAHB (orange and yellow lines) are also shown with the limits (blue dotted lines) used to select the stars. The middle and right panels show the scaled density of field stars along with the contaminants (red dots). The lower panels show the vector point diagram of the cluster and the two fields.}
\label{PM_arp2}
\end{center}
\end{figure*}

\begin{figure*}
\begin{center}
\includegraphics[width=\textwidth, height=\textheight, keepaspectratio]{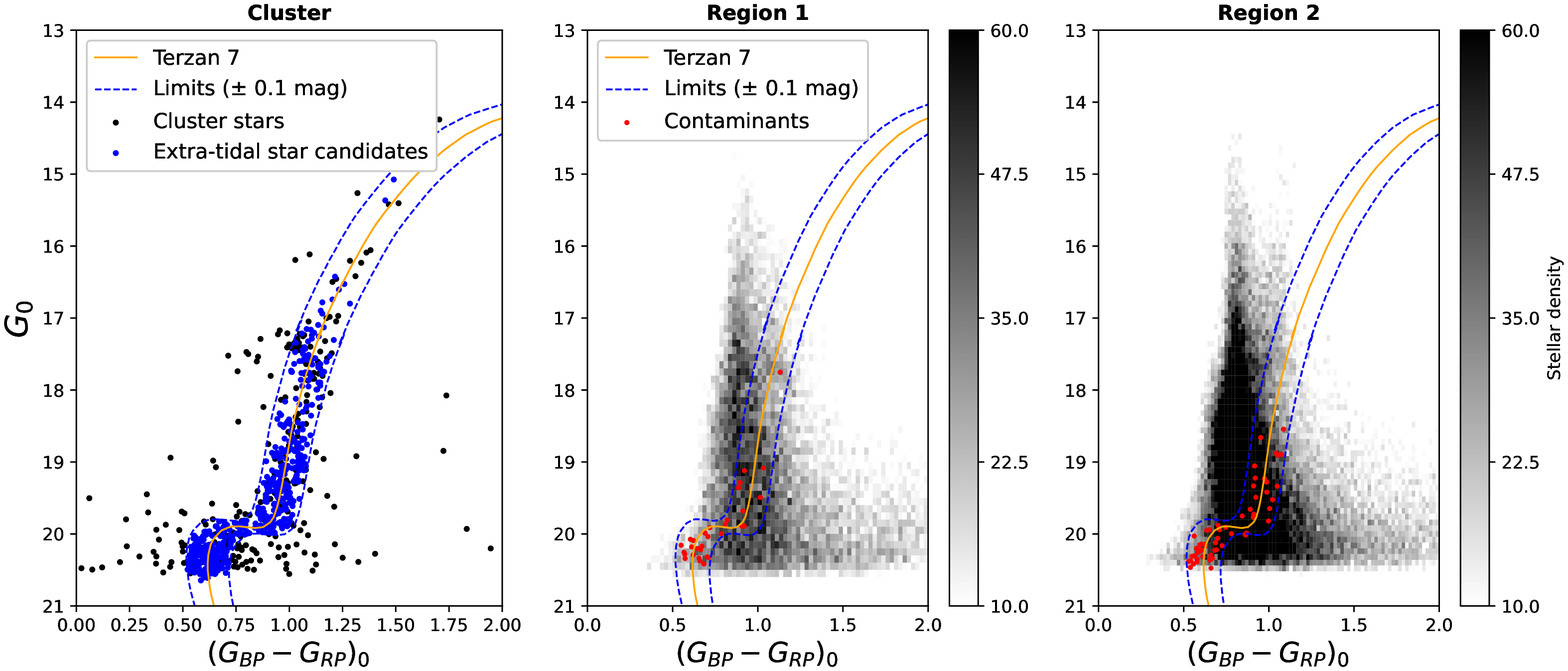}

\includegraphics[width=\textwidth, height=\textheight, keepaspectratio]{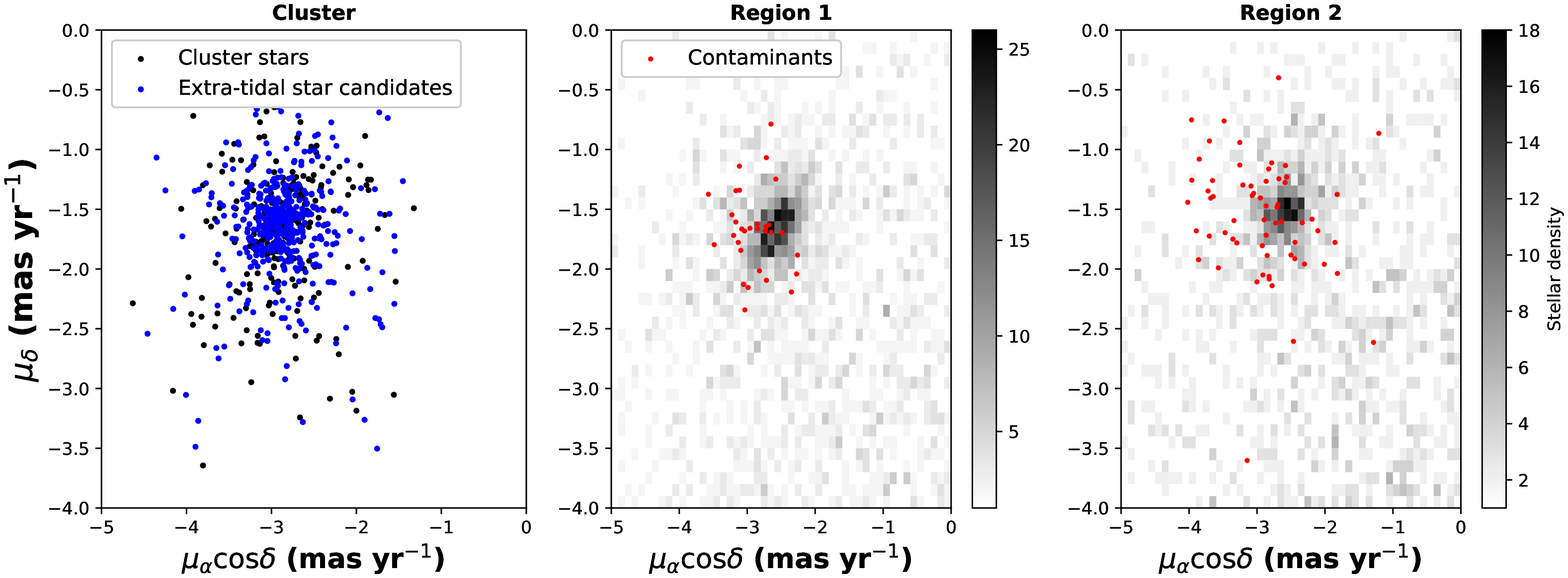}
\caption{Same as fig~\ref{PM_arp2} but for Terzan 7.}
\label{PM_terzan7}
\end{center}
\end{figure*}

\begin{figure*}
\begin{center}
\includegraphics[width=\textwidth, height=\textheight, keepaspectratio]{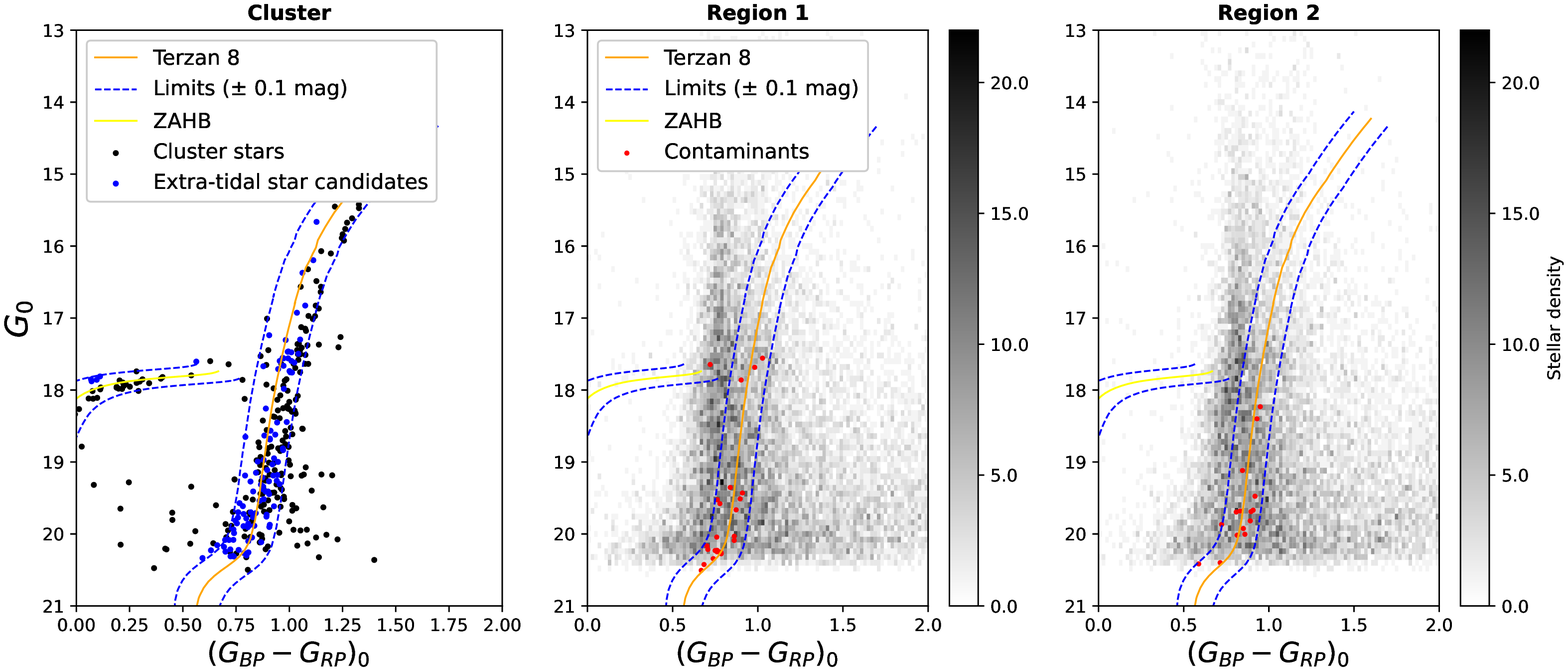}

\includegraphics[width=\textwidth, height=\textheight, keepaspectratio]{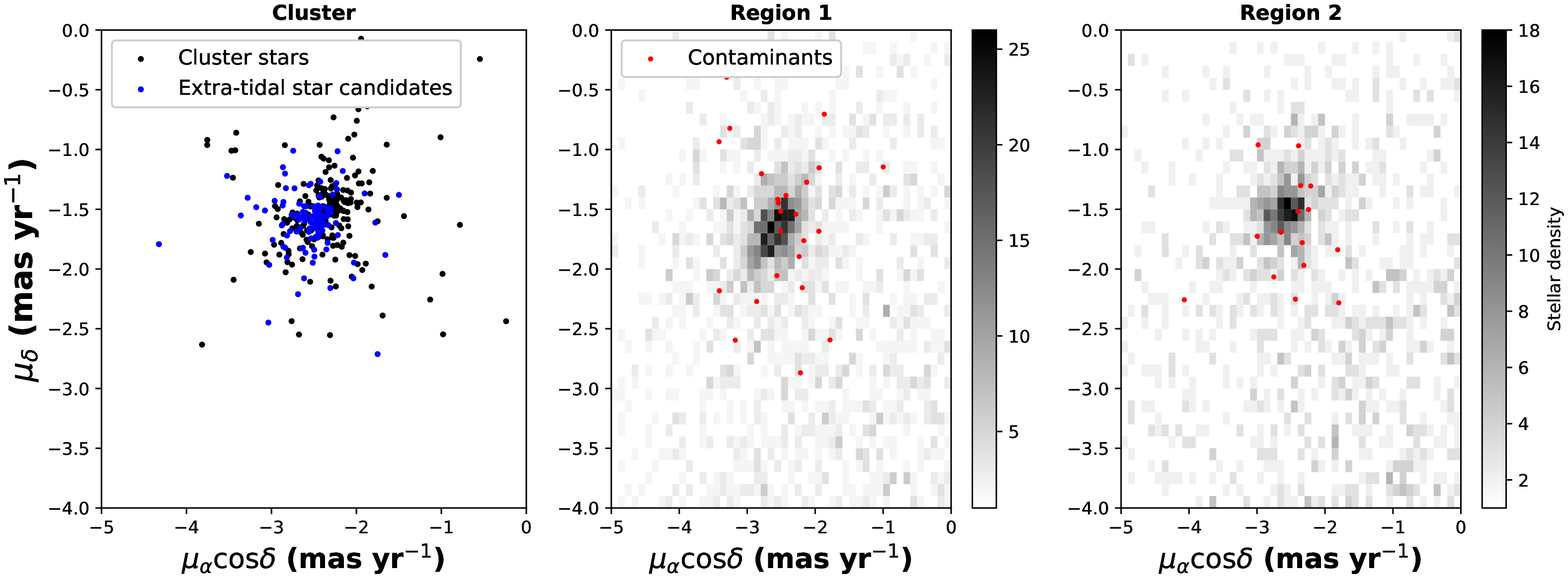}
\caption{Same as fig~\ref{PM_arp2} but for Terzan 8.}
\label{PM_terzan8}
\end{center}
\end{figure*}

In order to illustrate the accuracy of the estimates from our contamination analysis, we take the two template cases of the GCs Arp 2 and Terzan 7, which have different ages and metallicities that nicely bracket the whole range exhibited in the Sgr Age-Metallicity Relation. In this Appendix, we show the CMDs of two control fields (upper panels of Figure~\ref{PM_arp2}, Figure~\ref{PM_terzan7}, and  Figure~\ref{PM_terzan8}) located on both sides of the clusters (the specific location of these control fields do not alter the results). We also show the corresponding PM vector point diagrams (lower panels of Figure~\ref{PM_arp2}, Figure~\ref{PM_terzan7}, and  Figure~\ref{PM_terzan8}) for these fields. Evidently, there is more background in Arp 2 than in Terzan 7, mostly because Arp 2 is a larger cluster, but the contamination by field stars (bulge and Sgr) along these lines of sight is not a major problem. Table~\ref{tab:cont} lists the number of contaminants found around each control region, along with the scaled number of contaminants in that region.

\begin{table*}[htbp]
        \centering
        \caption{Number of contaminants selected in the control regions.}
        \label{tab:cont}
        \begin{tabular}{ccccc} 
\hline
Cluster ID & RA         &  DEC &        Number of stars &       Scaled number of stars \\
& (deg)         &  (deg) &       &       \\
\hline
Arp 2 & 293.732 & -33.314 & 459 & 64 \\
& 288.305 & -28.592 & 28 & 3 \\
Terzan 8 & 294.424 & -31.568 & 24 & 17 \\
& 293.041 & -35.064 & 15 & 11 \\
Terzan 7 & 293.041 & -35.064 & 62 & 88 \\
& 286.155 & -34.544 & 31 & 19 \\
NGC 5634 & 217.405 & -7.521 & 4 & 3 \\
& 217.405 & -4.431 & 3 & 2 \\
& 215.852 & -5.976 & 3 & 2 \\
& 218.958 & -5.976 & 3 & 2 \\
& 215.852 & -7.521 & 2 & 1 \\
& 215.852 & -4.431 & 2 & 1 \\
& 218.958 & -7.521 & 2 & 1 \\
& 218.958 & -4.431 & 3 & 2 \\
NGC 2419 & 114.537 & 37.507 & 7 & 6 \\
& 114.537 & 40.257 & 13 & 12 \\
& 112.771 & 38.882 & 12 & 9 \\
& 116.304 & 38.882 & 14 & 14 \\
& 112.771 & 37.507 & 11 & 8 \\
& 112.771 & 40.257 & 11 & 9 \\
& 116.304 & 37.507 & 4 & 3 \\
& 116.304 & 40.257 & 7 & 7 \\
NGC 4147 & 182.525 & 17.427 & 4 & 3 \\
& 182.525 & 19.658 & 3 & 2 \\
& 181.349 & 18.542 & 5 & 4 \\
& 183.701 & 18.542 & 1 & 0 \\
& 181.349 & 17.427 & 2 & 1 \\
& 181.349 & 19.658 & 2 & 1 \\
& 183.701 & 17.427 & 0 & 0 \\
& 183.701 & 19.658 & 6 & 5 \\
Pal 12 & 326.662 & -23.673 & 32 & 29 \\
& 326.662 & -18.833 & 16 & 14 \\
& 324.065 & -21.253 & 22 & 18 \\
& 329.259 & -21.253 & 25 & 25 \\
& 324.065 & -23.673 & 29 & 18 \\
& 324.065 & -18.833 & 14 & 11 \\
& 329.259 & -23.673 & 32 & 32 \\
& 329.259 & -18.833 & 15 & 15 \\
NGC 6284 & 256.121 & -25.441 & 59 & 39 \\
& 256.121 & -24.088 & 41 & 32 \\
& 255.376 & -24.764 & 44 & 38 \\
& 256.866 & -24.764 & 60 & 41 \\
& 255.376 & -25.441 & 40 & 31 \\
& 255.376 & -24.088 & 42 & 38 \\
& 256.866 & -25.441 & 68 & 46 \\
& 256.866 & -24.088 & 58 & 41 \\
\hline
\end{tabular}
\end{table*}
\end{appendix}

\end{document}